\documentclass[a4paper,11pt]{article}
\usepackage{graphicx}
\usepackage[space]{grffile}
\usepackage{latexsym}
\usepackage{textcomp}
\usepackage{longtable}
\usepackage{tabulary}
\usepackage{booktabs,array,multirow}
\usepackage{amsfonts,amsmath,amssymb}
\usepackage{natbib}
\usepackage{url}
\usepackage{hyperref}
\hypersetup{colorlinks=false,pdfborder={0 0 0}}
\usepackage{etoolbox}
\makeatletter
\usepackage{authblk}
\usepackage[utf8]{inputenc}
\usepackage[english]{babel}
\usepackage[top=1in, bottom=1in, left=1in, right=1in]{geometry}
\title{Dry and Moist Atmospheric Circulation with Uniform Sea-Surface Temperature}
\date{}
\author[1]{D.L. Suhas \thanks{Corresponding author: D.L. Suhas, suhasdl.mysore@gmail.com}}
\author[1,2]{Jai Sukhatme}
\author[3]{Nili Harnik}
\affil[1]{Centre for Atmospheric and Oceanic Sciences, Indian Institute of Science, Bangalore, 560012, India}
\affil[2]{Divecha Centre for Climate Change, Indian Institute of Science, Bangalore, 560012, India}
\affil[3]{Department of Geosciences, Tel Aviv University, Tel Aviv 69978, Israel}

\begin{document}

\maketitle

\begin{abstract}

The steady and transient response of "dynamically" dry and moist atmospheres to uniform sea-surface temperature (SST) is studied. 
Specifically, the latent heat ($L_v$) of water vapor is varied, so that for small $L_v$, water substance is essentially a passive tracer from a dynamical point of view. Despite the lack of SST gradients, a general circulation with Hadley and Ferrel cells is observed for relatively stronger moist coupling. Organized precipitation patterns via equatorial waves appear to play a significant role in tropical ascent, and along with the equatorial deformation radius, the Hadley cell width increases with coupling strength.
An abrupt switch to a much shallower tropical cell is noted when the system becomes completely passive. In all cases, the Hadley cell is thermally indirect and is influenced by eddy fluxes which are strong in the upper and lower troposphere.  Moist static energy is transported equatorward in the tropics and a larger amount is directed poleward in the midlatitudes. As a whole, there is an almost invariant poleward transport of moist static energy for relatively strong coupling of water substance.
Transient extratropical activity is seen in the form of intense warm-core vortices for strong coupling, and these systems become weaker and smaller as $L_v$ decreases. The drift of these moist vortices results in the observed poleward energy transport in the midlatitudes. 
In the tropics, intraseasonal variability is dominant and systematically shifts to longer time periods with stronger coupling. In fact, 
large-scale, low-frequency Kelvin waves and MJO-like modes disappear as water vapor becomes passive in nature.
Finally, extreme rainfall events associated with cyclonic storms vanish as water vapor becomes dynamically inactive, however, moderate precipitation events increase leading to higher total precipitation for weaker coupling of water substance. Tropospheric heating due to a saturation of the outgoing longwave radiation results in an increase in the stability of the atmosphere for strong coupling, and provides a plausible physical mechanism for interpreting the behavior of precipitation. \\

\end{abstract}

\section{Introduction}

\noindent
The explicit response of the atmosphere to uniform lower boundary forcing has proven to be a useful idealization in the climate modeling hierarchy \citep{maher2019model}. Simulations of this kind have yielded insight into possible routes to an organized general circulation \citep{sumi1992pattern,kirtman2000spontaneously,barsugli2005tropical,horinouchi2012moist} and the formation of tropical storm-like vortices and their sensitivity to environmental parameters on a $f$-plane \citep{held2008horizontally,khairoutdinov2013rotating,zhou2014parameter, ramirez2020spontaneous} as well as on the globe \citep{shi2014large,merlis2016surface,chavas2019dynamical}. A recent review of these efforts can be found in \citet{merlis2019aquaplanet}. Moist tropical transients, especially the Madden-Julian Oscillation (MJO) have also been studied in uniform sea-surface temperaure (SST) experiments \citep{grabowski2003mjo,grabowski2004moisture}. Further, constant SST experiments have allowed for an examination of convective self-aggregation in a variety of cloud resolving and parameterized general circulation models \citep[see, for example,][]{wing2017convective}. Another motivation comes from paleoclimate estimates that suggest a much weaker meridional gradient in SST during the early Pliocene \citep{brierley2009greatly}, and it is of interest to understand how such boundary conditions can affect the atmospheric circulation and transient phenomena, especially the frequency and strength of tropical and extratropical cyclones \citep{fedorov2010tropical,fedorov2019tropical}. 

\noindent Taking a steady state view, aquaplanet experiments with constant SST showed the development of an organized tropical atmospheric circulation due to the interaction of moist convection with rotation \citep{sumi1992pattern,kirtman2000spontaneously}. This took the form of a well defined 
equatorial convergence zone, along with tropical easterlies and subtropical westerlies \citep{kirtman2000spontaneously}. Further explorations along these lines showed that the nature of the tropical convergence zone was dependent on the magnitude of the uniform SST. In fact, single, double, symmetric and asymmetric convergence zones were seen to form under differing SST strengths \citep{barsugli2005tropical}. Interestingly, on varying the threshold relative humidity for triggering the deep moist convection, a uniform SST yielded tropical meridionally overturning cells with vastly differing strengths. In particular, low and high thresholds produced a weak and strong moist Hadley cells, respectively \citep{horinouchi2012moist}. 

\noindent Focusing on transient phenomena, uniform SST runs on a $f$--plane aimed at understanding rotating radiative-convective equilibrium, were seen to result in the spontaneous formation of tropical storm-like vortices \citep{held2008horizontally}. These results were expanded to larger domains \citep{zhou2014parameter}, and seen to carry over to a spherical geometry where multiple vortices were seen to form, persist and drift poleward \citep{shi2014large}. Moreover, the number of systems produced was seen to be dependent on the magnitude of SST \citep{khairoutdinov2013rotating,merlis2016surface}. The genesis of these storms, preferential latitudes and scales of the systems have also been examined in experiments with varying rates of planetary rotation \citep{chavas2019dynamical}. In the aforementioned simulations with varying thresholds for triggering deep convection, persistent cyclones along with organized tropical waves were observed in runs with high relative humidity threshold \citep{horinouchi2012moist}. At long times, after the establishment of zonal mean states, sporadic well organized eastward propagating convective activity was observed in the tropics \citep{sumi1992pattern}. In fact, detailed  simulations with cloud resolving convective parameterization have noted moisture-convection feedback and the emergence of modes resembling the MJO in flat SST experiments \citep{grabowski2003mjo,grabowski2004moisture}. 

\noindent From a self-aggregation perspective, uniform SST simulations have demonstrated the spontaneous appearance of organized moist convection and this was attributed to the feedbacks involving cloud-radiative interactions \citep{bretherton2005energy}. The sensitivity of aggregation to the resolution of models, the domain size and magnitude of SST have been examined and linked to the differing distribution of clouds in these scenarios \citep{muller2012detailed}. Further, the initiation of aggregation and its maintenance were seen to be primarily driven by the shortwave and longwave radiative processes \citep{wing2014physical}. 
The self-aggregation of moisture is believed to play a vital role in the genesis of tropical cyclones and its intensification \citep{shi2014large,muller2018acceleration}, as well as in the emergence of MJO \citep{arnold2015global}.

\noindent Here we adopt the viewpoint of treating water vapor as a dynamically active scalar field \citep{sobel2002water}. In particular, we note that the coupling of water vapor, or any other condensable substance, is controlled by its latent heat $(L_v$). When $L_v \rightarrow 0$, the condensable substance is advected with the flow but will not be dynamically coupled to the equations of motion\footnote{The so called advection-condensation framework sets $L_v=0$, and has proven useful in probing the distribution of water vapor in the troposphere and its radiative impacts \citep{pierrehumbert2007relative,ogorman2006stochastic,sukhatme2011advection}}. An interesting question to ask in this uniform SST scenario is, what happens when water vapor is no longer (or only weakly) dynamically active? For example, does the interaction of convection and rotation still result in an organized meridional flow? Do we observe a tropical Hadley cell? Is there a systematic transport of energy across latitudes even without a SST gradient, and if so, why? And how does the partition of latent and dry static energy change in these simulations? What is the nature of the transient tropical wave activity? Specifically, given that the coupling with water vapor is thought to be essential to its existence \citep{grabowski2003mjo,raymond2009moisture,sobel2013moisture,adames2016mjo}, do we observe an intraseasonal mode like the MJO when water vapor behaves more like a passive tracer? Further, if there is a change in tropical intraseasonal activity, is it abrupt? How does the nature of the midlatitude synoptic vorticity field change as one progresses from a dynamically dry to moist atmosphere. Do we always see the formation of tropical storm-like vortices? How about the nature of precipitation and cloud fraction in a world with small $L_v$? 

\noindent In all, we believe that these simulations, much like the moist and dry scenarios studied by \citet{frierson2006gray,frierson2007gray}, will help in a more robust view of planetary atmospheric circulation regimes in the presence of a condensable substance, especially ones with a different latent heat compared to water vapor. Further, even on present-day Earth, our hope is that exploring the effect of different coupling strengths will help in the understanding of moist geophysical systems --- much like the dry and moist simulations of tropical cyclones that are yielding insight into the fundamental nature of these cyclonic systems \citep{Ag,Chav_dry,wang_lin}. The modeling framework to address these questions is described in Section 2. The results pertaining to the mean flow and transient activity are presented in Sections 3 and 4, respectively. Section 5 collects and discusses the main findings of our investigation.


\section{Modeling Framework}

The aquaplanet experiments are carried out using the Community Atmosphere Model, Version 5.3
(CAM 5.3), the atmospheric component of the Community Earth System Model (CESM), Version 1.2.2.
The model uses a finite volume dynamical core with a resolution of 0.9$^{\circ}$ latitude $\times$ 1.25$^{\circ}$ longitude, with 30 vertical levels. Based on the recommendations of \citet{medeiros2016reference}, aerosol effects are minimized by removing the aerosol emissions and by specifying constant droplet and ice number concentrations in the microphysics. Zhang-McFarlane convection scheme is used for deep convection, while the University of Washington (UWSC) scheme is employed for shallow convection.
Following \citet{shi2014large}, we have used the aquaplanet configuration with a prescribed globally uniform SST of 27$^{\circ}$C. Uniform solar insolation is used with no diurnal cycle \citep{barsugli2005tropical,kirtman2000spontaneously} and the solar constant was set to 342 $\mathrm{W/m^2}$. All other fields (like Ozone) were set to their annual horizontally averaged values. In essence, except the Coriolis force, no other meridional gradients are present. 

\noindent To mimic atmospheres with differing strengths of moist coupling, we vary the latent heat of vaporization $L_v$. 
Specifically, latent heat is varied everywhere except in the Clausius-Clapeyron equation as detailed in \citet{frierson2006gray}. 
The degree of moist coupling can also be controlled by varying the amount of moisture in the atmosphere (by varying the saturation vapor pressure at a reference temperature) and keeping $L_v$ fixed \citep{frierson2006gray}, but we prefer varying $L_v$ due to the ease in modifying this parameter in the model and the fact that this directly controls the amount of heat released on condensation.  Further, changing $L_v$ can potentially be useful in understanding other planetary atmospheres, such as Mars \citep{Readbook} or Titan \citep{Mitchell-titan}, that have condensable substances with very different latent heats than water vapor. Specifically, as $L_v \rightarrow 0$, water substance is transported by the flow and is active in the radiation budget,  but does not lead to latent heating in the dynamical equations of motion. Note that for $L_v=0$ run, we also explicitly switch off the evaporation in the model. When $L_v > 0$, water substance becomes dynamically active and its strength, or coupling with the flow increases with $L_v$. 
Perhaps due to excessive amount of heat released, the model fails at latent heats around twice the latent heat of water. So we restrict our experiments  to  $L_v$ between 1.5X to 0X, where X =  $2.5 \times 10^6$ J/kg. 
The model is run for 3 years, with the first year of data being discarded as a spin-up time.

\section{Zonal Mean Picture}

\begin{figure}
    \centering
    \includegraphics[width=\textwidth]{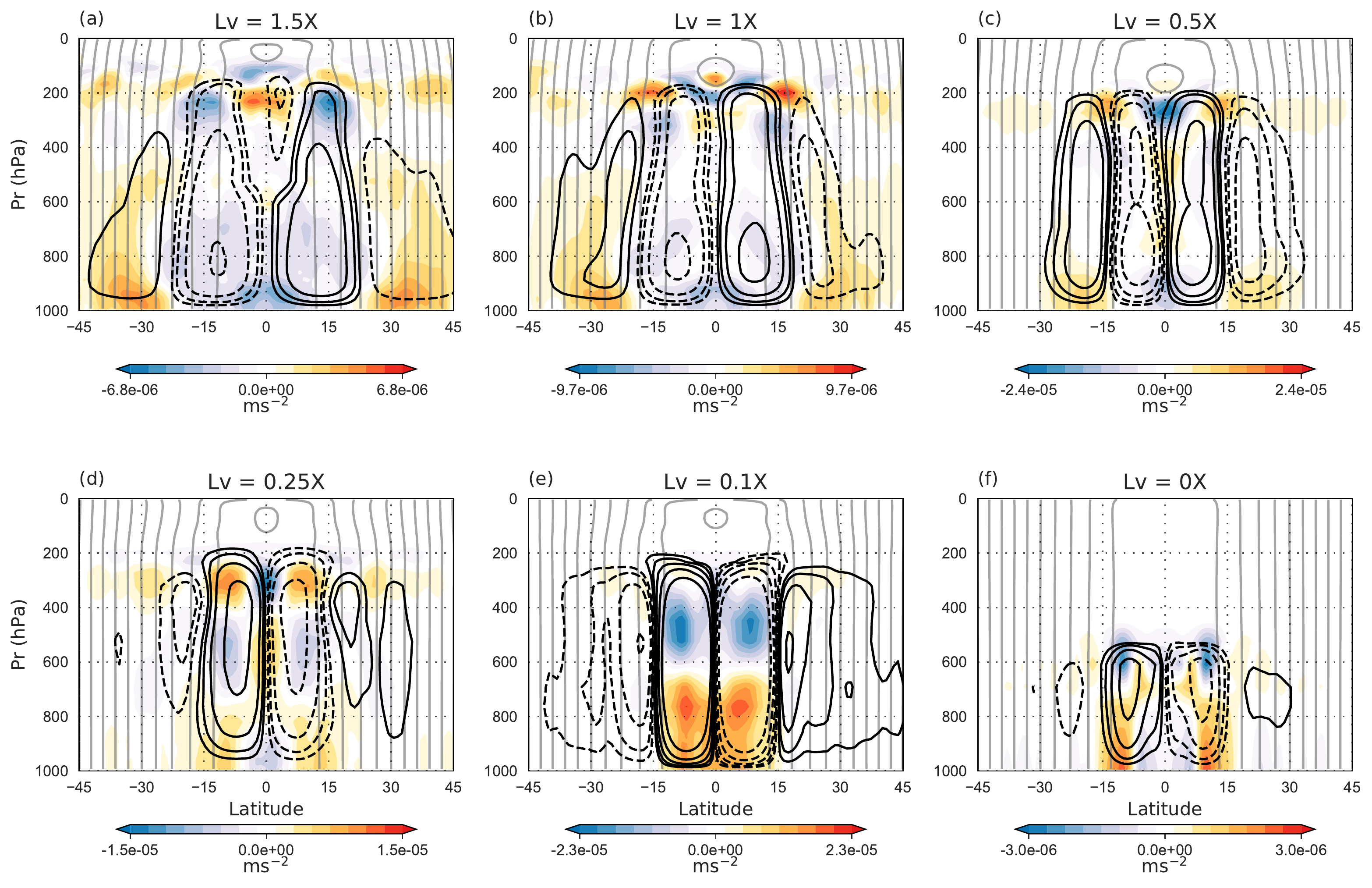}
    \caption{Mean meridional mass stream function (black contours; solid clockwise and dashed counterclockwise) and angular momentum (grey contours) averaged over the last two years of the runs. The divergence of eddy momentum flux is shown in color. 
    The stream function contours are logarithmic in nature,  with its magnitude doubling between successive contours. 
    The lowest contour has a value of $2.5 \times 10^9$ kg s$^{-1}$ and zero contour is not shown. Angular momentum contours are in the intervals of $0.1 \Omega a^2$, with decreasing values away from the equator.}
    \label{fig:AquaHadley}
\end{figure}

\noindent Even with no solar and surface meridional thermal gradients imposed, a systematic Hadley cell like circulation emerges in all the cases. This is seen in the mean meridional mass streamfunction presented in Figure \ref{fig:AquaHadley} as a pair of tropical cells straddling the equator. In addition, weaker oppositely directed Ferrel-like circulation cells are seen in the midlatitudes. For $L_v =$ 1X, the meridional streamfunction is of the same order of magnitude, but narrower than that of the annual mean circulation of present-day Earth \citep{dima2003seasonality,walker2005response}. 
Usually, the annual mean value is less than the seasonal extremes, but in these simulations, there are no seasons and this may explain the comparable magnitude even without a SST gradient. In all these cases, i.e., from $L_v =$ 1.5X to 0X, the vorticity based local Rossby number at the extremities of the Hadley cell is between $(0.3,0.6)$ --- thus, idealized theories from limiting cases of Rossby number $\rightarrow 0$ or $1$ will not immediately be applicable \citep{walker2006eddy}.
The strength of these cells compares well with the results obtained by \citet{horinouchi2012moist}, although \citet{kirtman2000spontaneously} obtained much weaker cells with a uniform SST forcing. 
Further, \citet{horinouchi2012moist} observed a weakening of Hadley cell circulation with a lowering of the threshold for moist convection trigger, but we find no significant changes in magnitude with varying $L_v$. As water vapor becomes passive, the tropical cells become narrower (Figure \ref{fig:AquaHadley}c), while the Ferrel cells seem more clearly formed. In fact, there is a systematic decrease in the Hadley cell width from about 25$^\circ$ to 15$^\circ$ as $L_v$ varies from 1.5X to 0X. Remarkably, for $L_v \le$ 0.25X, the mean tropical circulation reverses with poleward surface transport in the tropics. The sense of the mean meridional cells in the midlatitudes also reverses. 
A similar reversal in the atmospheric circulation was observed by \citet{barsugli2005tropical} with variations in insolation and uniform SSTs. 
The limiting case of a dry atmosphere, shown in Figure \ref{fig:AquaHadley}f, indicates a much shallower tropical cell. 
Comparing panels e and f of Figure \ref{fig:AquaHadley}, the dramatic change in the height of the tropical circulation suggests that the dynamically active nature of moist processes appears to act as a trigger or switch in sustaining the deep convection. 

\noindent The zonal mean angular momentum contours in Figure \ref{fig:AquaHadley} are nearly vertical, and flow streamlines cross these contours in each of the runs, indicating a strong influence of eddies in shaping this circulation \citep{walker2006eddy,schneider2008eddy}. The divergence of the resolved eddy angular momentum flux, $\nabla \cdot$ [cos$\phi$ ($\overline{u'v'}$, $\overline{u'\omega'}$)], is shown by colored shading in Figure \ref{fig:AquaHadley}. Unlike eddy fluxes in the present-day atmosphere \citep{ait2015eddy}, the vertical component of momentum flux convergence is also significant, especially as water vapor becomes passive in nature. Also, these fluxes are dominant in both the upper troposphere and near the surface of the planet. In fact, these flat SST cases constitute an intermediate regime between present-day Earth where the eddy momentum fluxes are concentrated in the upper troposphere, and reverse surface temperature gradient scenarios on Earth as well as high obliquity planets where eddy fluxes are confined near the surface \citep{ait2015eddy,kang2019tropical}. Interestingly, the resolved eddy fluxes seem to account well for the crossing of angular momentum surfaces by the overturning circulation in the lower troposphere, but not so well in the upper troposphere, where their structure is more complex than the overturning circulation. This implies the existence of another momentum sink, possibly unresolved gravity waves which arise near the tropopause, or a numerical sink of momentum \citep{tonn}. Further, as $L_v$ becomes smaller, the eddy momentum flux is progressively restricted to the tropical regions.

\begin{figure}
    \centering
    \includegraphics[width=\textwidth]{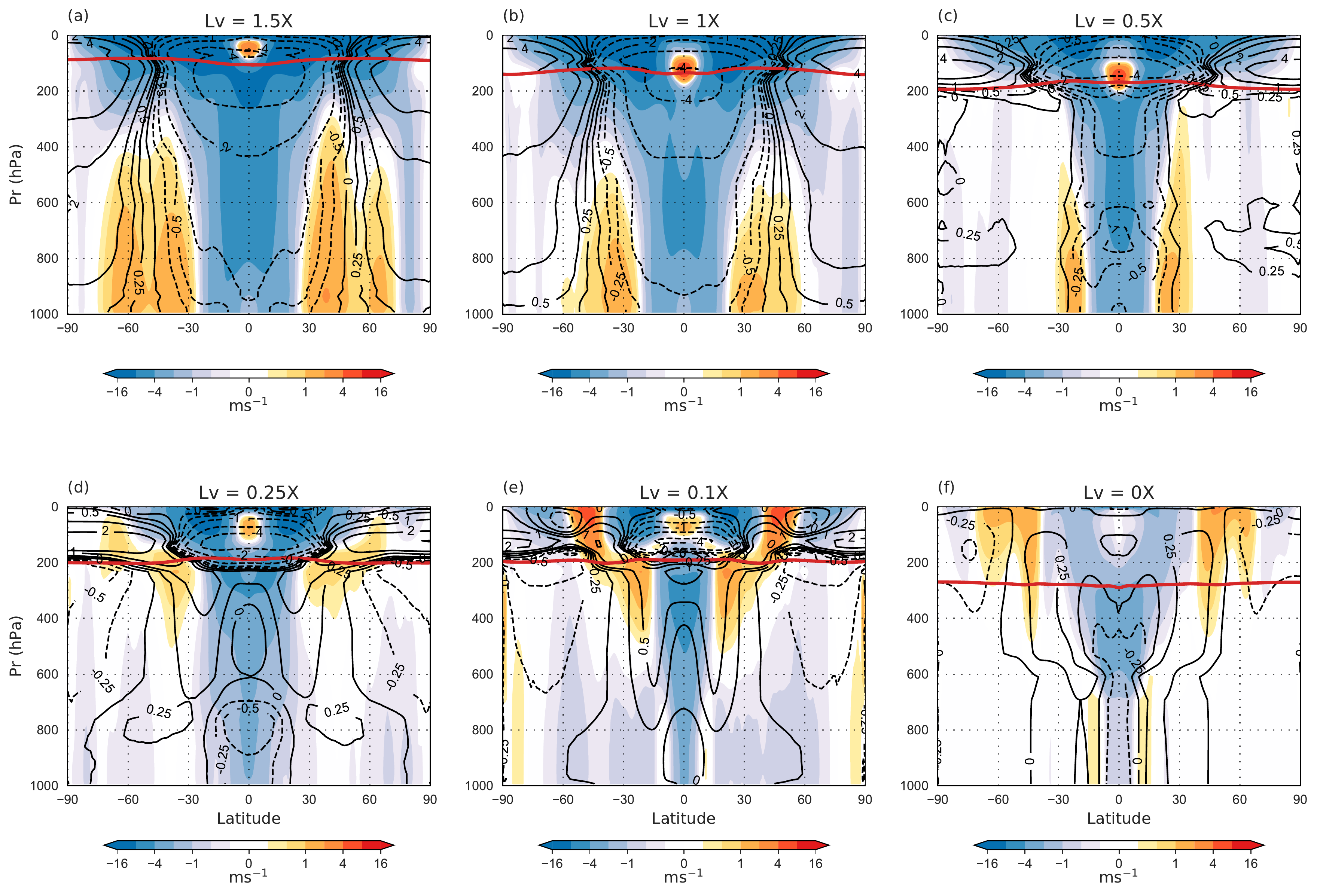}
    \caption{Zonal mean zonal wind (in color) along with zonal temperature anomalies (black contours) averaged over the last two years of the run. Temperature anomalies are computed by removing the mean at each pressure level.  The contour levels are logarithmic in nature with its magnitude doubling between successive contours. Tropopause is shown as a thick red line.}
    \label{fig:AquaUwindTemp}
\end{figure}

\begin{figure}
    \centering
    \includegraphics[width=\textwidth]{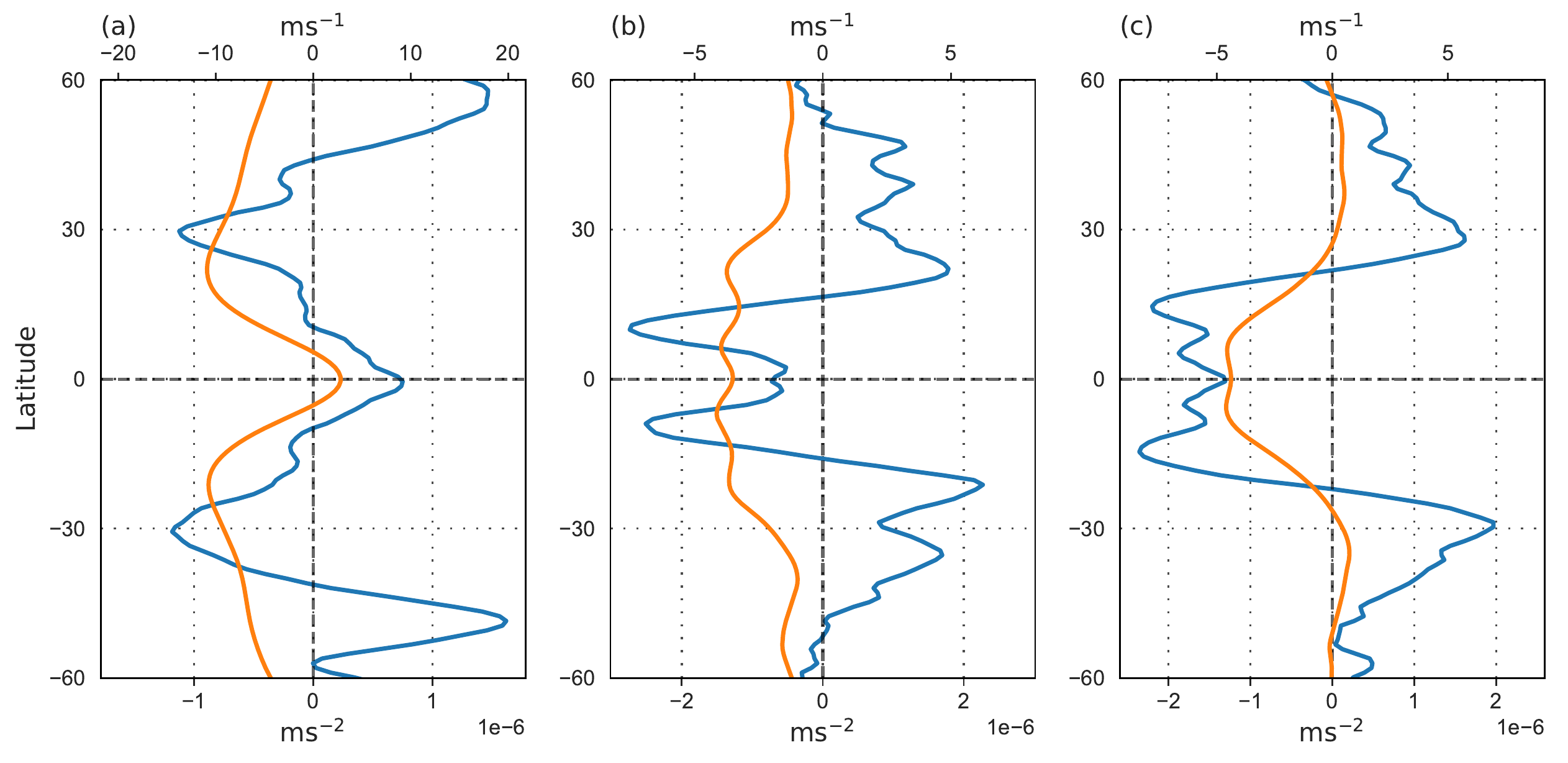}
    \caption{Zonal momentum flux convergence (blue) and zonal mean zonal wind (orange) for $L_v=$ 1X. The plots are averaged over (a) 50-150 hPa, (b) 200-400 hPa and (c) 500-800 hPa respectively, and is computed using the last two years of the run.  }
    \label{fig:AquaZonalMomtFlux}
\end{figure}

\noindent The zonal mean zonal winds as a function of height and
latitude are shown in Figure \ref{fig:AquaUwindTemp}. For all the cases, easterlies form through most of the tropical troposphere and a small region of superrotation is seen directly above the equator near the tropopause. Indeed, low pole to equator SST gradients are known to favour tropical superrotation \citep{lutsko2018response}. The magnitude of superrotation becomes weaker, moves into the stratosphere and progressively disappears as $L_v \
\rightarrow 0$. 
For $L_v =$ 1.5X, 1X and 0.5X cases, westerlies form in the subtropics, though mostly in the lower and middle levels.
Whereas for lower latent heats, westerlies are pushed equatorward and upwards between 200-400 hPa. 
The momentum flux convergence that is responsible for driving the zonal mean zonal flow is shown in Figure \ref{fig:AquaZonalMomtFlux}. 
From a flux decomposition \citep{lee1999climatological}, we note that transient eddies play a dominant role in the momentum flux budget.
This is expected because there is no orography, land-sea contrast or imposed SST structure, hence stationary eddies are absent. Further, the mean meridional circulation also does not have a seasonal cycle and its contribution to the momentum flux convergence is small \citep{lee1999climatological,dima2005tropical}. As is illustrated for $L_v=$ 1X in Figure \ref{fig:AquaZonalMomtFlux}, the eddy flux convergence drives the upper equatorial superrotation (Figure \ref{fig:AquaZonalMomtFlux}a), the equatorial easterlies below 200 hPa (Figure \ref{fig:AquaZonalMomtFlux}b) and the westerlies in the midlatitudes from 30-60$^\circ$ below 500 hPa (Figure \ref{fig:AquaZonalMomtFlux}c). Even in other cases, the eddy convergence seems to be primarily responsible for driving the zonal mean flow. In the midlatitudes, we also see a transition from lower level westerly jets in the Ferrel-like cell regions of the strong moisture coupled runs, to narrow upper-tropospheric and stratospheric westerly jets for the weaker moisture coupled runs. We note that the change from a surface to an upper tropospheric westerly jet goes along with a change in the direction of the overturning cells. In fact, the sharp westerly jets between 200-400 hPa at around 15$^\circ$-20$^\circ$ latitude in the $L_v=$ 0.1X runs are maintained by a vertical advection of zonal momentum by the strong ascent at around 600 hPa (Fig. 1e).
The existence of easterlies at all latitudes for the low moisture coupled runs implies a net gain of atmospheric angular momentum from the surface. This net gain of momentum is actually found in all runs (the globally averaged surface zonal winds are on average negative for all runs), and is consistent with the model not conserving momentum \citep{tonn}.

\noindent The zonal mean temperature anomalies, i.e., the deviation with respect to the mean on a given pressure level, is also shown in Figure \ref{fig:AquaUwindTemp}. Due to the absence of any prescribed horizontal thermal gradients, the temperature anomalies are much weaker than the typical Earth-like case. While the anomalies are less than 0.5$^\circ$C in the lower troposphere, the largest anomalies are seen in the upper atmosphere with magnitudes reaching up to  4$^\circ$C. 
For $L_v \ge $0.5X, the equator is slightly colder than the poles, a feature that was also observed in some of the earlier studies with uniform SSTs \citep{kirtman2000spontaneously, barsugli2005tropical, shi2014large}. 
Interestingly, for weaker coupling of water vapor, the situation changes and the equatorial region is marginally warmer than the poles. Note that this change in the equator to pole temperature gradient  is accompanied by a reversal in the sense of tropical circulation. 
Thus, in all cases, the tropical cell is an indirect circulation. In fact, the criterion for a moist direct circulation \citep{emanuel1995thermally} is almost never satisfied by these simulations, and as noted above, the coherent circulation cells in the troposphere are strongly influenced by the eddy fluxes. With a lack of imposed temperature gradients at the surface, the sense of temperature anomalies in the upper troposphere in Figure \ref{fig:AquaUwindTemp} is consistent with the overturning circulation and its reversal seen in Figure \ref{fig:AquaHadley}. Further, the direction of zonal mean flow in the upper midlatitudes is also in accord with the meridional temperature gradient in the troposphere.

\begin{figure}
    \centering
    \includegraphics[width=\textwidth]{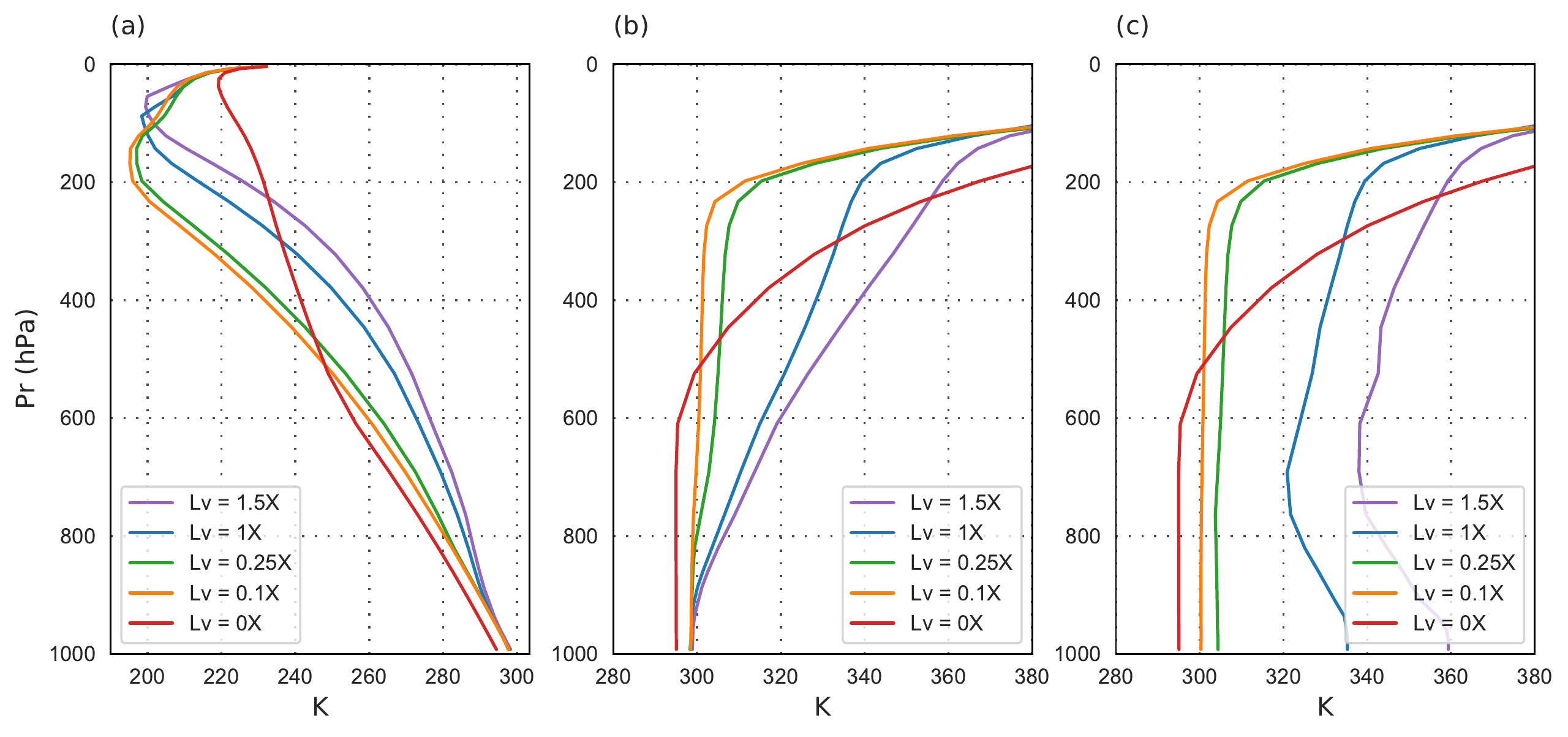}
    \caption{Vertical profile of (a) temperature, (b) potential temperature and (c) equivalent potential temperature in the tropics (30$^{\circ}$N/S). The profile is averaged over the last two years of the run.}
    \label{fig:AquaTempVerticalProfile}
\end{figure}

\noindent Vertical profiles of temperature, potential temperature and equivalent potential temperature are shown in Figure \ref{fig:AquaTempVerticalProfile}. 
The temperature lapse rate increases with decreasing moist coupling, with the $L_v = $ 0X approaching the dry lapse rate.
Potential temperature increases with height and in general, lower latent heat cases have lower potential temperature.
Equivalent potential temperature is mostly constant till up to 200 hPa, especially for low coupling cases of $L_v=$ 0.1X and 0.25X. For the more active cases, $L_v=$ 1.5X and 1X, it decreases till about 700 hPa and then rises gradually up to a height of 200 hPa. As expected, with decreasing moist coupling the equivalent potential temperature and potential temperature profile converges.
The decrease of equivalent potential temperature with height (but not the potential temperature) indicates that the conditions are unstable for moist ascent but stable for dry ascent in the lower troposphere. Further, higher up in the troposphere, we note an increase in stability with $L_v$. This is reminiscent of the present-day tropics \citep{holton2012book}, but note that, here the vertical profiles shown in Figure \ref{fig:AquaTempVerticalProfile} are similar in the tropical and midlatitudinal regions. Thus, the conditional instability observed for strong coupling is present in the tropics as well as at higher latitudes. There is a sharp change in potential temperature at upper levels (Figure \ref{fig:AquaTempVerticalProfile}b) roughly coinciding with the level of tropopause (shown by red line in Figure \ref{fig:AquaUwindTemp}). But, for the dry case, which has a constant potential temperature up to 600 hPa and rising rapidly above it, the identified tropopause level is significantly higher. This is due to our usage of the WMO criterion of 2 K/km lapse rate to identify the level of tropopause. As seen from Figure \ref{fig:AquaTempVerticalProfile}a, for the dry case, even though the temperature lapse rate decreases above 600 hPa, its not sufficiently low to identify as the tropopause by the WMO criterion, and hence the discrepancy. But, from a dynamical point of view, the tropical overturning circulation and eddy fluxes for the dry case are confined below 600 hPa.

\begin{figure}
    \centering
    \includegraphics[width=\textwidth]{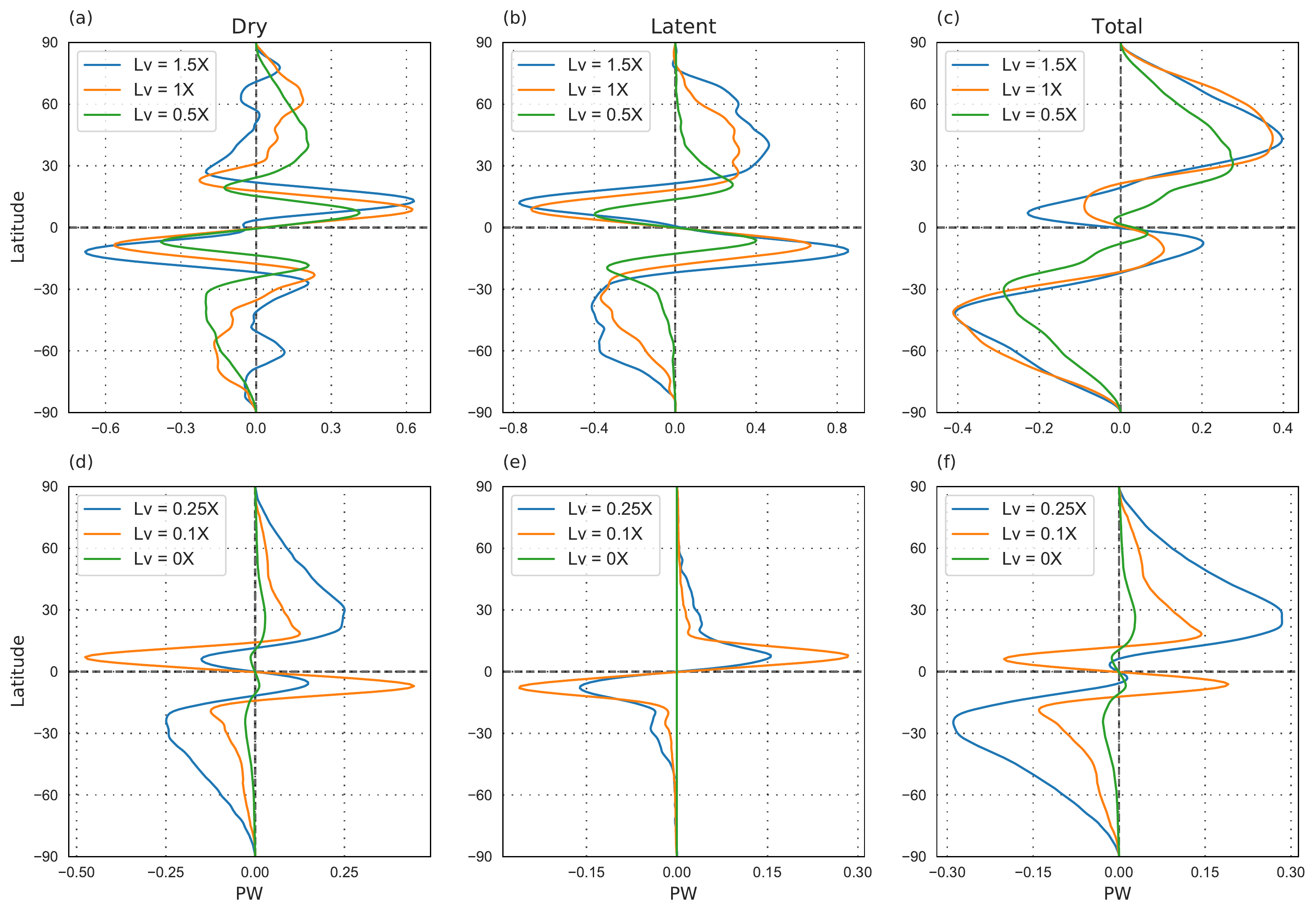}
    \caption{Vertically integrated dry, latent heat and moist static energy fluxes averaged over the last two years of the run.}
    \label{fig:AquaMseFlux}
\end{figure}

\begin{figure}
    \centering
    \includegraphics[width=\textwidth]{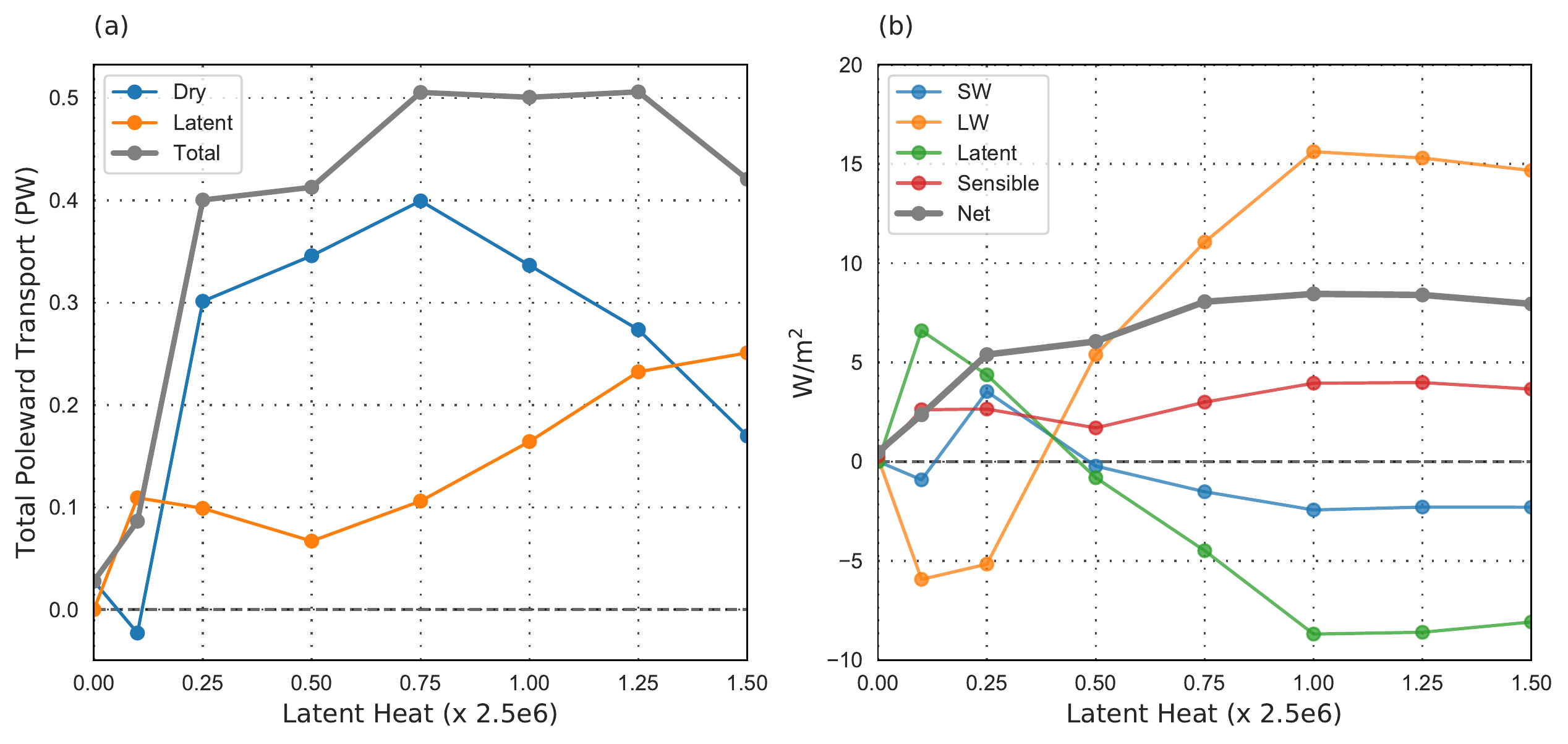}
    \caption{(a) Global mean poleward transport of energy as a function of latent heat. (b) Meridional gradient of various flux components (positive sign implying a net flux into the atmosphere). Here, gradient is taken as the difference between the averaged values in the tropics (30$^\circ$ N/S) and extra-tropics (30$^\circ$ - 90$^\circ$ latitudes). The plots are computed using the last two years of the run.}
    \label{fig:AquaMsePolewardTransportFluxGradient}
\end{figure}

\begin{figure}
    \centering
    \includegraphics[width=\textwidth]{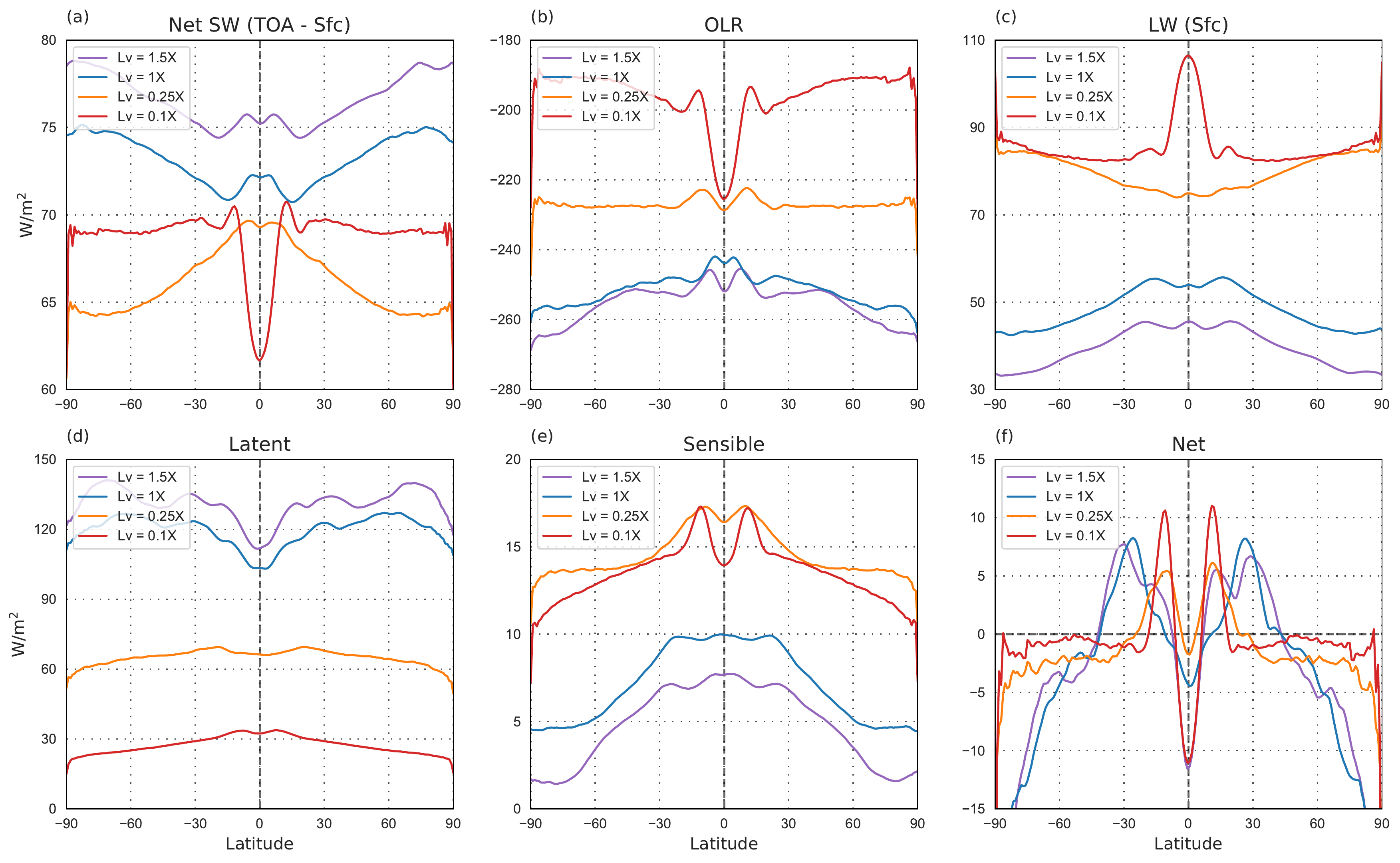}
    \caption{Zonal mean time mean fluxes for four different simulations; specifically, $L_v=$ 1.5X, 1X, 0.25X and 0.1X. The various components are, (a) net atmospheric absorption of shortwave radiation (top - surface fluxes), (b) longwave radiation at the top (OLR), (c) longwave radiation at the surface, (d) latent heat, (e) sensible heat and (f) net flux into the atmosphere. The fluxes are averaged over the last two years of the run and positive sign implies a net flux into the atmosphere.}
    \label{fig:AquaNetFlux}
\end{figure}

\noindent The vertically integrated dry and moist static energy fluxes are shown in Figure \ref{fig:AquaMseFlux}.
Moist static energy (MSE), $m = C_pT + gz + L_vq$, comprises of the dry static energy $ C_pT + gz$ and latent heat energy $L_vq$. Vertically integrated moist static energy flux is then defined as $2\pi a cos\phi \int_{0}^{p_s} \overline{vm} dP/g$, where $\phi$ is the latitude, $a$ is the radius of earth, $g$ is the acceleration due to gravity, $p_s$ is the surface pressure and overbar denotes a time and zonal mean \citep{frierson2007gray}. Tropical latent heat flux is equatorward (poleward) for higher (lower) latent heats, i.e., it mimics the sense of mean circulation at the surface. The dry static energy flux opposes the latent term in the tropics; thus, MSE transport is determined by the difference of these two opposing components. For the more active cases, i.e., $L_v >$ 0.25X, the latent term is larger, while for $L_v\le$ 0.25X, the dry static term contributes more to the MSE. But, in all cases, net MSE transport is equatorward in the tropics and is dominated by the mean flow. The situation is quite different in the midlatitudes. Here, almost all the cases show poleward latent heat and dry static energy transports, and thus a robust poleward MSE transport. Further, transport in the midlatitudes is almost completely accounted for by transient eddies. Of course, transport by latent heat generally decreases with the moist coupling, and for $L_v\le$ 0.25X it almost vanishes in the midlatitudes. The MSE poleward transport in the midlatitudes is greater than the equatorward flow in the tropics. As a result, despite the flat SSTs, there is a net transport of MSE toward the poles. This is the case for all $L_v$ values, regardless of the direction of the zonal mean MSE gradients (which are similar to the temperature anomalies, Figure \ref{fig:AquaUwindTemp}). Thus, while for the weakly coupled runs the midlatitude MSE fluxes are down gradient, for the strongly coupled runs, the midlatitude MSE fluxes are up gradient, suggesting the process driving the flux is not a simple eddy diffusivity or baroclinic instability. 
Remarkably the magnitude of total MSE transport (Figure \ref{fig:AquaMsePolewardTransportFluxGradient}) remains roughly the same for strong coupling (till $L_v =$ 0.25X) and falls drastically as water substance becomes almost passive. The nearly constant transport is due to an increase of dry static energy transport that compensates the decreasing latent heat component with $L_v$. This is reminiscent of a similar compensation found by \citet{frierson2007gray}, when moisture content is varied but with a meridional surface temperature gradient. Similar to the mean meridional circulation, the meridional energy transport is also an order smaller than the present-day Earth like conditions. 
Further, a decomposition of moist static energy flux into mean and eddy components suggests that the major contribution is from transient eddies in the midlatitudes, while the mean flow dominates the transport in the tropics. In all, as with the appearance of a coherent circulation, despite the lack of an imposed SST gradient, the atmospheric flow systematically transports energy towards the poles and this is almost invariant for cases with relatively strong moisture coupling. 

\noindent A net poleward MSE flux implies a net input of energy into the atmosphere at low latitudes, and a net output of energy at high latitudes \citep{Tren_step}. 
We examine how the net Energy Input into the Atmosphere (EIA), and it's latitudinal gradient changes with the strength of moisture coupling. Figure \ref{fig:AquaNetFlux} shows the latitudinal profile of the time and zonal mean latent, sensible and net longwave radiation at the surface, as well as the outgoing longwave radiation (OLR) at the top of the atmosphere and the net atmospheric absorption of shortwave radiation for a few $L_v$ values. We see that as $L_v$ increases, the energy exchange at the surface, which acts to heat the atmosphere, changes from being dominated by longwave radiation to being dominated by latent heat (LH) fluxes. Consistently, the longwave radiation seems to saturate at low $L_v$ values (values similar for $L_v=$ 0.1X and $L_v=$ 0.25X, c.f. Figure \ref{fig:AquaNetFlux}c), while LH starts to saturate at large $L_v$ (only a small increase between $L_v=$ 1X and $L_v=$ 1.5X, Figure \ref{fig:AquaNetFlux}d). We also note an opposing dependence of LH and longwave radiation on the atmospheric temperature -- warmer atmospheric temperatures increase LH due to the larger moisture holding capacity of the air, but they decrease the net longwave cooling of the surface because of the larger downward flux from the atmosphere (while SST is fixed). As a result, the mean latitudinal gradient of these fields (Figure \ref{fig:AquaMsePolewardTransportFluxGradient}b) changes following the temperature field -- as $L_v$ increases, the mean latitudinal gradient of net surface longwave radiation increases, while that of LH decreases (directly via the increase in $L_v$ and indirectly through the temperature effect)
\footnote{Note that for clarity of the presentation, Figure \ref{fig:AquaMsePolewardTransportFluxGradient}b shows the mean gradient of the total longwave radiation i.e., net upward flux at surface minus OLR. While these fluxes have an opposing dependence on atmospheric temperature, the variation or their mean latitudinal gradient  with $L_v$ is similar for values $L_v\ge$ 0.5X, and is quite noisy below, due to the changes of the tropical overturning circulation.}. 
The sensible heat (SH) fluxes have a relatively small contribution to the overall surface energy budget, however, their contribution to the mean latitudinal gradient is significant (Figure \ref{fig:AquaNetFlux}e), with values being strongest in the tropics for all runs, but the tropics--high latitudes gradient is largest for large $L_v$. This is probably due to two factors affecting the sensible heat fluxes-- the strength of surface winds and the atmospheric temperatures. For large $L_v$, with warmer poles, the colder temperatures and stronger surface winds make the equatorial SH fluxes larger than at the poles, but for the small $L_v$ runs, the warmer tropics act to reduce the equator--pole gradient. The other atmospheric heating term is the net shortwave radiation absorption. This quantity depends mostly on the clouds, which change in a complex way (especially for small $L_v$), but its overall contribution peaks at high latitudes for large $L_v$, while for small $L_v$ the latitudinal gradient reverses, and changes strongly especially in the tropics for $L_v=$ 0.1X. The above heating terms are balanced by the OLR, which increases with $L_v$, and saturates at around $L_v=$ 1X. This increase is consistent with the decrease in temperature lapse rate as $L_v$ increases, since the longwave emitted to space originates in the mid-upper troposphere \citep{ray-book}. At smallest $L_v$, OLR has a strong peak in the tropics, and a flat extratropical profile, but this changes as $L_v$ is increased to emissions peaking at high latitudes for large $L_v$. Summing all these contributions to get the net EIA (Figure \ref{fig:AquaNetFlux}f), we see that the atmosphere gains energy in the subtropics and emits energy in the extratropics and near the equator. This is consistent with the subtropical peak of MSE flux for small $L_v$ and the stronger midlatitude peaks of MSE flux for large $L_v$ (Figure \ref{fig:AquaMseFlux}), and is of course quite different from present-day Earth with differential solar heating. Despite the somewhat noisy latitudinal profile of EIA, we see a clear tendency for the latitudinal gradient to increase with $L_v$, so that for large $L_v$, the high latitudes cool most strongly and the subtropics heat most strongly. This change in high minus low latitude EIA (Figure \ref{fig:AquaMsePolewardTransportFluxGradient}b) is consistent with the increase in mean poleward MSE flux as $L_v$ increases (Fig \ref{fig:AquaMsePolewardTransportFluxGradient}a). 

\noindent It is again interesting to point out that for strong moisture coupling, the MSE fluxes are upgradient, i.e., from the colder tropics to the warmer poles.   Consistently, we will see later on that the poleward MSE flux is driven by the dynamics of vortices on a sphere, resulting in the high latitudes being warmer than the tropics.
At small $L_v$ values, the anomalies are more wave-like, and the EIA is dominated by radiative processes (OLR and SW absorption) to yield a net positive heating in the tropics. The wave-like anomalies, in this case, act diffusively and flux MSE poleward, reducing (but not reversing) the temperature gradient. 


\section{Transient Disturbances}

\begin{figure}
    \centering
    \includegraphics[width=0.5\textwidth]{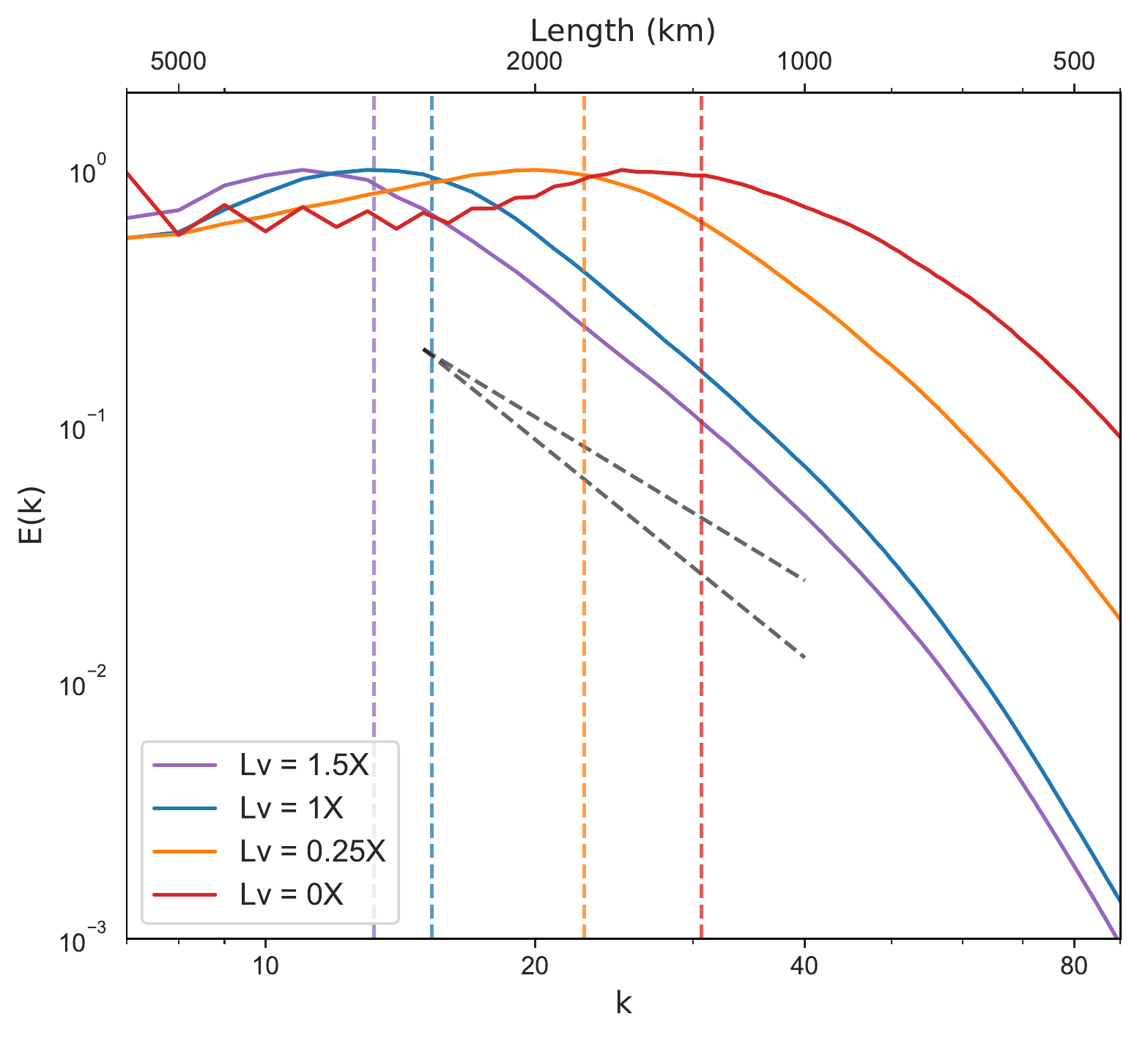}
    \caption{Vertically integrated kinetic energy spectra averaged over the last two years of the run. The spectra is integrated from the surface till 200 hPa and is normalized for easy comparison. Black dashed lines have slopes of $-3$ and $-4$. Deformation radius is denoted by dashed vertical lines.}
    \label{fig:AquaKeSpectra}
\end{figure}

\noindent In the previous section, we looked at the emergent steady state zonal mean circulation and its maintenance by diabatic processes and by eddy MSE and angular momentum fluxes. Given that there is no differential heating imposed in these runs, the underlying processes driving these eddy fluxes are not apriori clear. In this section we examine the characteristics of the transient eddies and vortices in more detail.
The globally integrated kinetic energy (KE) spectra for various scenarios is shown in Figure \ref{fig:AquaKeSpectra}. The KE dominates over the available potential energy \citep{shi2014large} and further, the rotational part of the spectrum has much more energy than the divergent component, so Figure \ref{fig:AquaKeSpectra} is essentially a plot of the rotational kinetic energy. 
Signs of power-law scaling emerge with stronger moist coupling, and the exponent of the spectrum tends to be around $-3$ to $-4$ at intermediate scales and falls off steeply at smaller scales. This exponent is consistent with the known scaling of rotational modes in idealized models of rotating stratified turbulence \citep[see, for example,][]{bartello1995geostrophic,kitamura2006kh,sukhatme2008vortical,vallgren2011possible} as well as from midlatitude upper tropospheric \citep{nastrom-gage,lindborg} and ocean surface \citep{BCF,rocha,sukhatme2020} observations. Further, the scaling persists up to about the deformation scale  \citep{shi2014large}. With weakening moist coupling, the deformation scale reduces with a corresponding shift in the peak of energy spectra towards the smaller scales (Figure \ref{fig:AquaKeSpectra}). Also, the spectrum becomes ``shallower" with decreasing latent heat. In fact, for the passive run, the KE spectrum is more appropriately characterized as a plateau, rather than a power-law, with maximum variance near the deformation scale.

\begin{figure}
    \centering
    \includegraphics[width=0.8\textwidth]{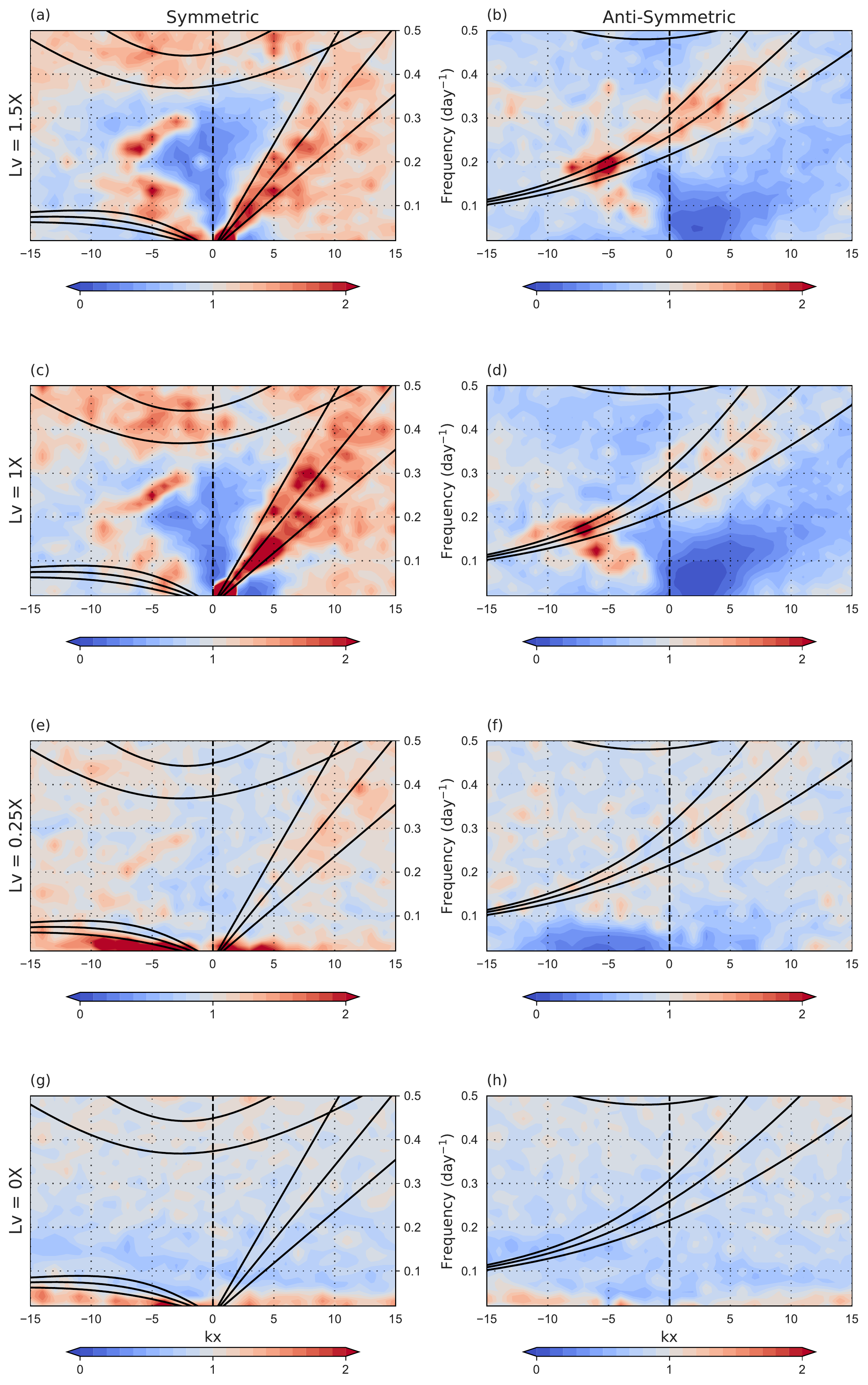}
    \caption{Frequency--wavenumber power spectra of zonal wind at 850 hPa. Symmetric and anti-symmetric components are shown in the first and second column, respectively and the background spectra is removed. The power spectra is calculated over the last two years of the run and is averaged over 15$^\circ$N to 15$^\circ$S latitudes. Superimposed are the dispersion curves (in black)  with an equivalent depth of 12, 25 and 50m. } 
    \label{fig:AquaWk}
\end{figure}

\noindent Next, we examine transient activity in the tropics and midlatitudes. 
With regard to the tropics, frequency--wavenumber plots of low-level horizontal winds are shown in Figure \ref{fig:AquaWk}. 
The symmetric and anti-symmetric components are separated and the background spectra is removed. With stronger moist coupling, most of the variability in the symmetric component lies along the linear Kelvin and Rossby wave dispersion curves with Earth-like wave speeds, and much like in other aquaplanet studies \citep{shi2014large,das2016low,shi2018wishe}, there is an eastward propagating MJO-like signature at low frequencies and large scales. The anti-symmetric part of the spectrum in Figure \ref{fig:AquaWk} shows the presence of mixed Rossby-gravity waves. 
Further, activity in the ``tropical depression zone" \citep{wheeler1999convectively} is also evident in the stronger coupling runs. It is important to note that these wavenumber--frequency diagrams represent scales larger than those accounted for in the spectra of Figure \ref{fig:AquaKeSpectra}. The intraseasonal modes, especially the eastward moving MJO and Kelvin waves disappear with decreasing latent heat, i.e., as water substance becomes dynamically passive. 
The lack of a MJO-like mode for weak coupling suggests that this mode of low frequency activity --- at least in this modeling framework --- requires interactive moisture \citep{sobel2013moisture, adames2016mjo}. 

\noindent The loss of Kelvin waves is evident by the $L_v=$ 0.25X case itself. In fact, together with evidence on the disappearance of Kelvin modes with changes in parameters like the relative humidity threshold for triggering convection \citep{horinouchi2012moist}, convective time scales \citep{frierson2007convectively} and background saturation fields \citep{suhas2020}, this points to the sensitivity of these waves in the tropical atmosphere. For the limiting case of $L_v=$ 0X (Figure \ref{fig:AquaWk}g,h), it appears as though there isn't much variability in the tropics at these large length scales. 

\begin{figure}
    \centering
    \includegraphics[width=\textwidth]{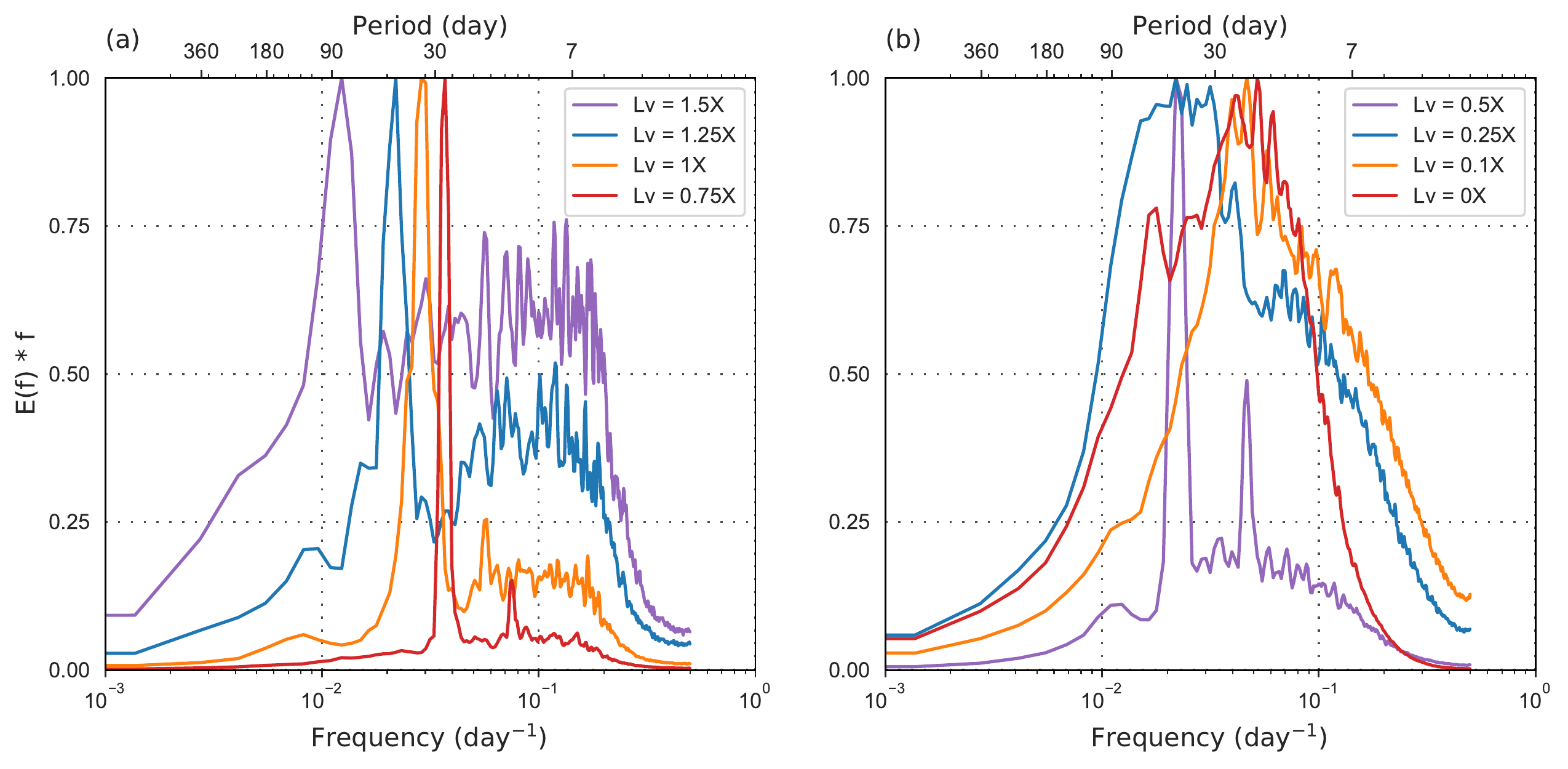}
    \caption{Variance preserving spectra of zonal wind at 850 hPa averaged over the latitudes 15$^\circ$N to 15$^\circ$S. The spectra is normalized  and is computed using the last two years of the run. } 
    \label{fig:AquaFreqVarSpectra}
\end{figure}

\begin{figure}
    \centering
    \includegraphics[width=\textwidth]{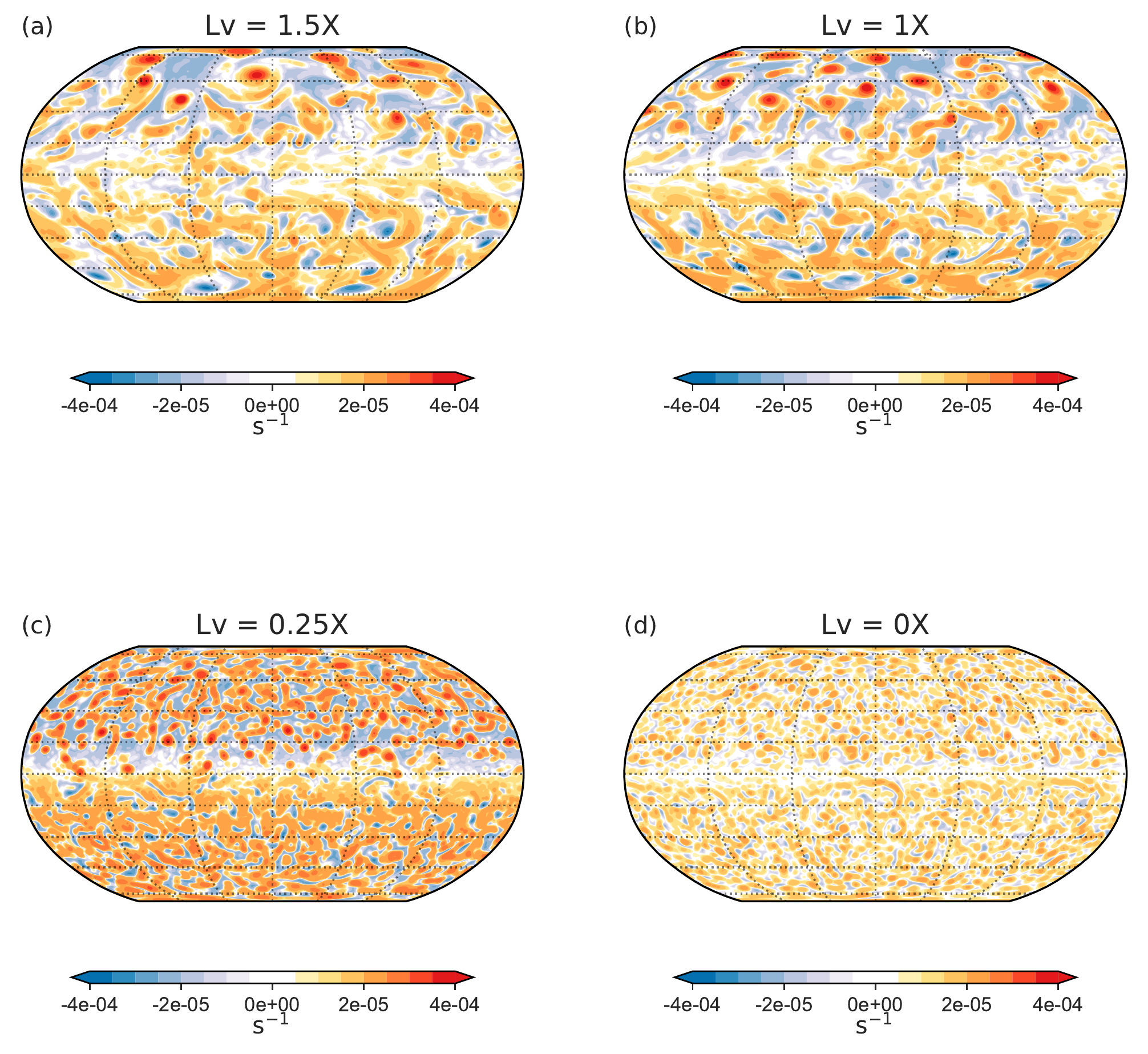}
    \caption{Snapshot of relative vorticity at 850 hPa on Day 1000 for four different simulations; specifically, $L_v=$ 1.5X, 1X, 0.25X and 0X. The contour levels are logarithmic in nature with its magnitude doubling between successive contours.} 
    \label{fig:AquaVorticity}
\end{figure}

\noindent In all simulations, intraseasonal variability (around 20--90 days) is dominant in the tropics. Specifically, tropical zonal winds in lower atmosphere are averaged to produce a daily time series whose spectral properties are shown in Figure \ref{fig:AquaFreqVarSpectra}. 
A sharp distinct peak is observed with stronger coupling, and the dominant period of activity shifts to longer time scales with increasing latent heats (Figure \ref{fig:AquaFreqVarSpectra}a), i.e., as $L_v=$ 0.75X $\rightarrow$ 1.5X, we note a peak that shifts from about 30 days to 90 days. In addition, for $L_v=$ 1.25X and 1.5X there is a second window of activity that is at shorter time scales, approximately 7 to 15 days.  However, with weakened coupling, especially for the cases with $L_v =$ 0.25X and below, rather than distinct peaks, the winds show more of a plateau that is spread around the period of a month (Figure \ref{fig:AquaFreqVarSpectra}b). Taken together, a broad picture that emerges in the tropics is that of progressively longer period intraseasonal variability that aligns with familiar tropical dispersion curves and co-exists with smaller scale (predominantly rotational) turbulence for stronger moist coupling. 

\begin{figure}
    \centering
    \includegraphics[width=\textwidth]{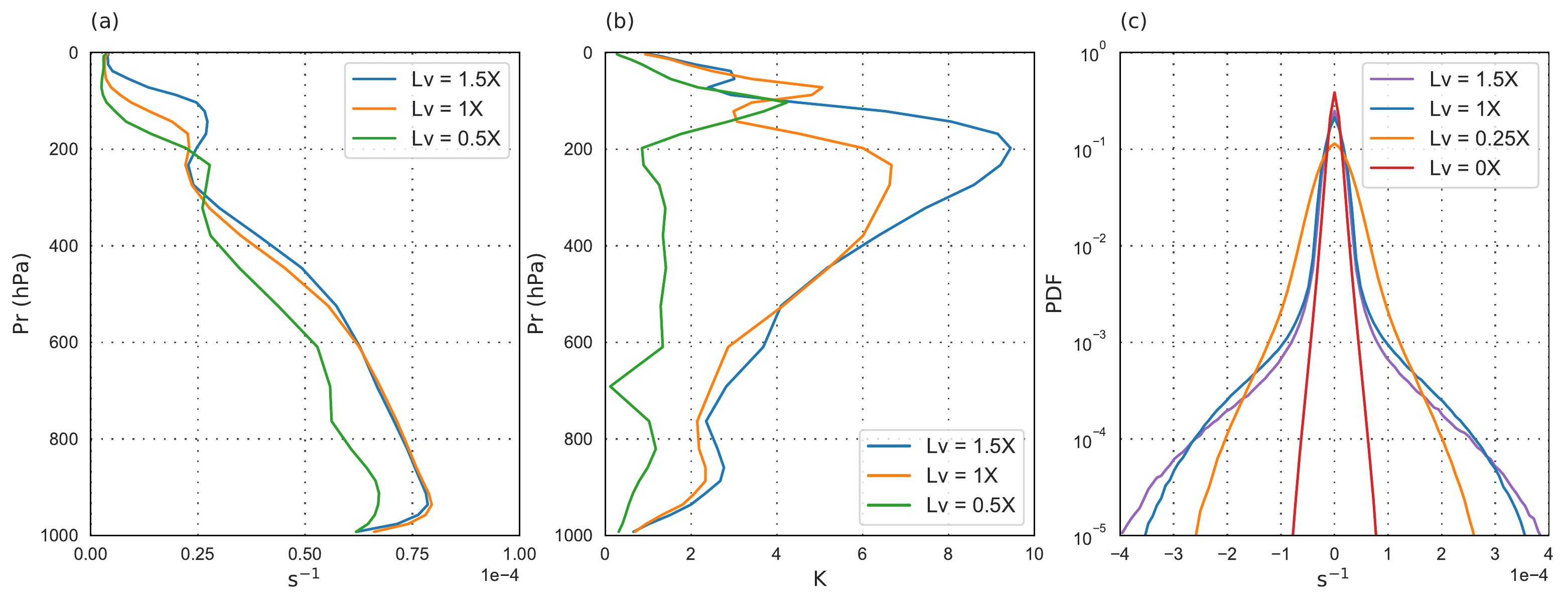}
    \caption{Vertical profile of (a) relative vorticity and (b) temperature anomaly inside the cyclonic structure. The profiles are computed by averaging over multiple cyclonic structures at Day 1000. Temperature anomaly is defined as the deviation from the global average at each vertical level. 
    Panel (c) shows the probability density function of relative vorticity field at 850 hPa, and is computed over the last two years of the run.}
    \label{fig:AquaVrtProfilePdf}
\end{figure}

\begin{figure}
    \centering
    \includegraphics[width=\textwidth]{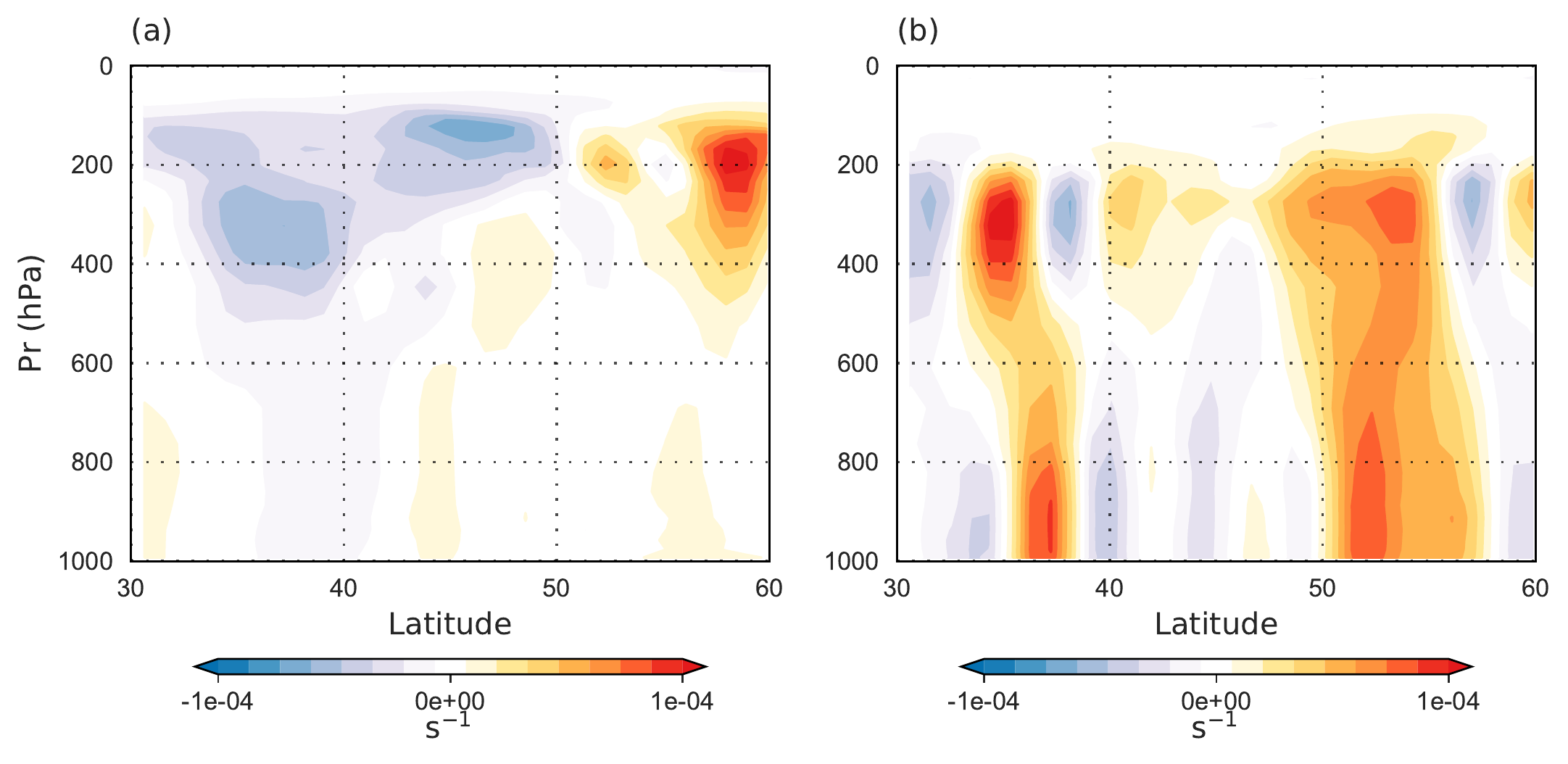}
    \caption{Vertical section of vorticity anomalies for (a) $L_v= 1X$ and (b) $L_v= 0.25X$ cases at Day 1000. Vorticity anomalies are the deviation from the vertical mean and is shown for 30$^\circ$N - 60$^\circ$N latitudes and 180$^\circ$ longitude. }
    \label{fig:AquaVrtVertical}
\end{figure}

\noindent With regard to transients at higher latitudes, a snapshot of the lower level relative vorticity (850 hPa) for Day 1000 is shown in Figure \ref{fig:AquaVorticity}. Much like prior observations in $f$--plane \citep{held2008horizontally} and global \citep{shi2014large,chavas2019dynamical} uniform SST simulations, intense cyclonic structures can be seen in the extra-tropics for relatively strong coupling. 
Animations of the vorticity field suggest that these storm-like vortices are born in the subtropics, sometimes merge to form larger structures and progress westward and towards the poles, due to $\beta$--drift \citep{shi2014large}. In fact, these storms are responsible for the poleward transport of both dry and latent energy as noted in Figure \ref{fig:AquaMseFlux}. The lifetime of these systems is approximately of the order of a month.
The vertical profile of a typical storm for different $L_v$ values is shown in Figure \ref{fig:AquaVrtProfilePdf}a, b. Here, the vorticity anomaly has a maximum in the lower troposphere and the system has a warm core with largest temperature anomalies in the upper troposphere. Thus, even though these storms are observed at higher latitudes, they are broadly similar to present-day tropical cyclones \citep{wang201913} and some tropical lows \citep{kushwaha}. 
Indeed, for $L_v=$ 1X, the largest temperature anomaly is at approximately 300 hPa and has a value of about 6 K, both aspects being comparable to present-day category one tropical cyclones \citep{wang201913}.  
The fact that such cyclones are supported at higher latitudes in these simulations is likely tied to the vertical thermal structure of the atmosphere, especially its similarity with tropical regions (Figure \ref{fig:AquaTempVerticalProfile}).
The strength of these vortices, the magnitude of temperature anomalies and their spatial scale decreases with lower latent heats and in the dry simulation (Figure \ref{fig:AquaVrtProfilePdf} and Figure \ref{fig:AquaVorticity}c and d). A similar pattern emerged in simulations of cyclones with increasing surface dryness \citep{Chav_dry}. This change in morphology is starkly captured in the probability density function (PDF) of the 850 hPa relative vorticity field as shown in Figure \ref{fig:AquaVrtProfilePdf}c. Quite clearly, for strong coupling the PDF is fat-tailed with a high propensity of extremes associated with intense cyclones, whereas for progressively smaller coupling the PDF tends to have exponential tails (seen as straight lines in the semi-log plot of Figure \ref{fig:AquaVrtProfilePdf}c). In addition to these deep tropospheric vortices, the midlatitudes also show shallow near tropopause waves. Examples of these systems can been seen in a snapshot of vorticity anomalies for a few $L_v$ cases in Figure \ref{fig:AquaVrtVertical}. Specifically, in this snapshot, for $L_v=$ 1X, we observe near tropopause anomalies, while for $L_v=$ 0.25X we capture both the near tropopause structures as well as the deeper poleward drifting vortices. Indeed, these disturbances are identified as waves by noting that they are associated with a systematic meridional heat transport (albeit much smaller than the MSE transport; Figure \ref{fig:AquaMseFlux}) of the form $\overline{v'T'}$. In fact, the near tropopause waves are tied to the relatively large temperature anomalies noted in Figure \ref{fig:AquaUwindTemp}.


\subsection{Nature of Precipitation}

\begin{figure}
    \centering
    \includegraphics[width=\textwidth]{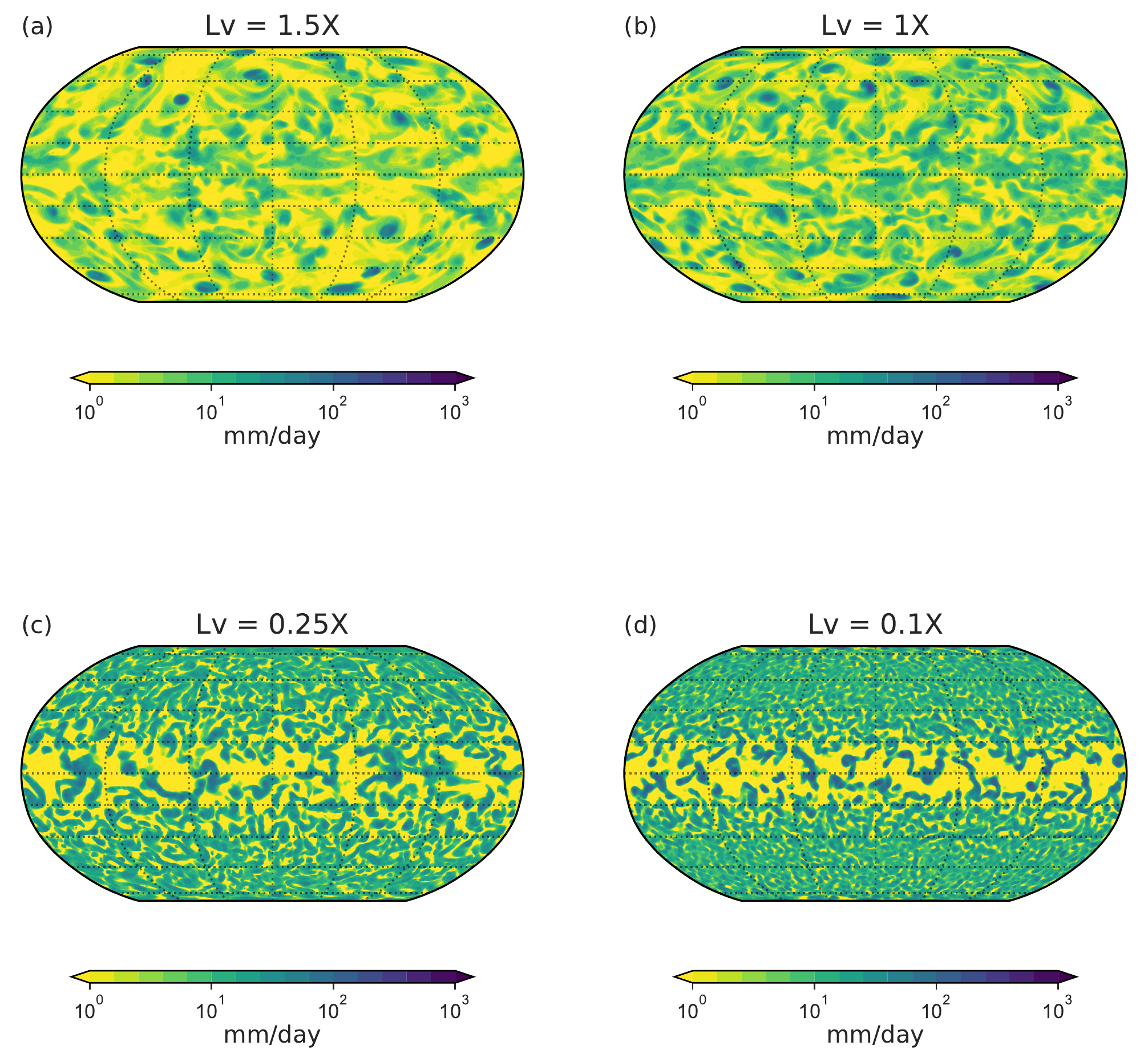}
    \caption{Snapshots of the daily precipitation at Day 1000 for four different simulations; specifically, $L_v=$ 1.5X, 1X, 0.25X and 0.1X. Note that the colorbar is logarithmic in nature.}
    \label{fig:AquaPrecipitation}
\end{figure}

\begin{figure}
    \centering
    \includegraphics[width=\textwidth]{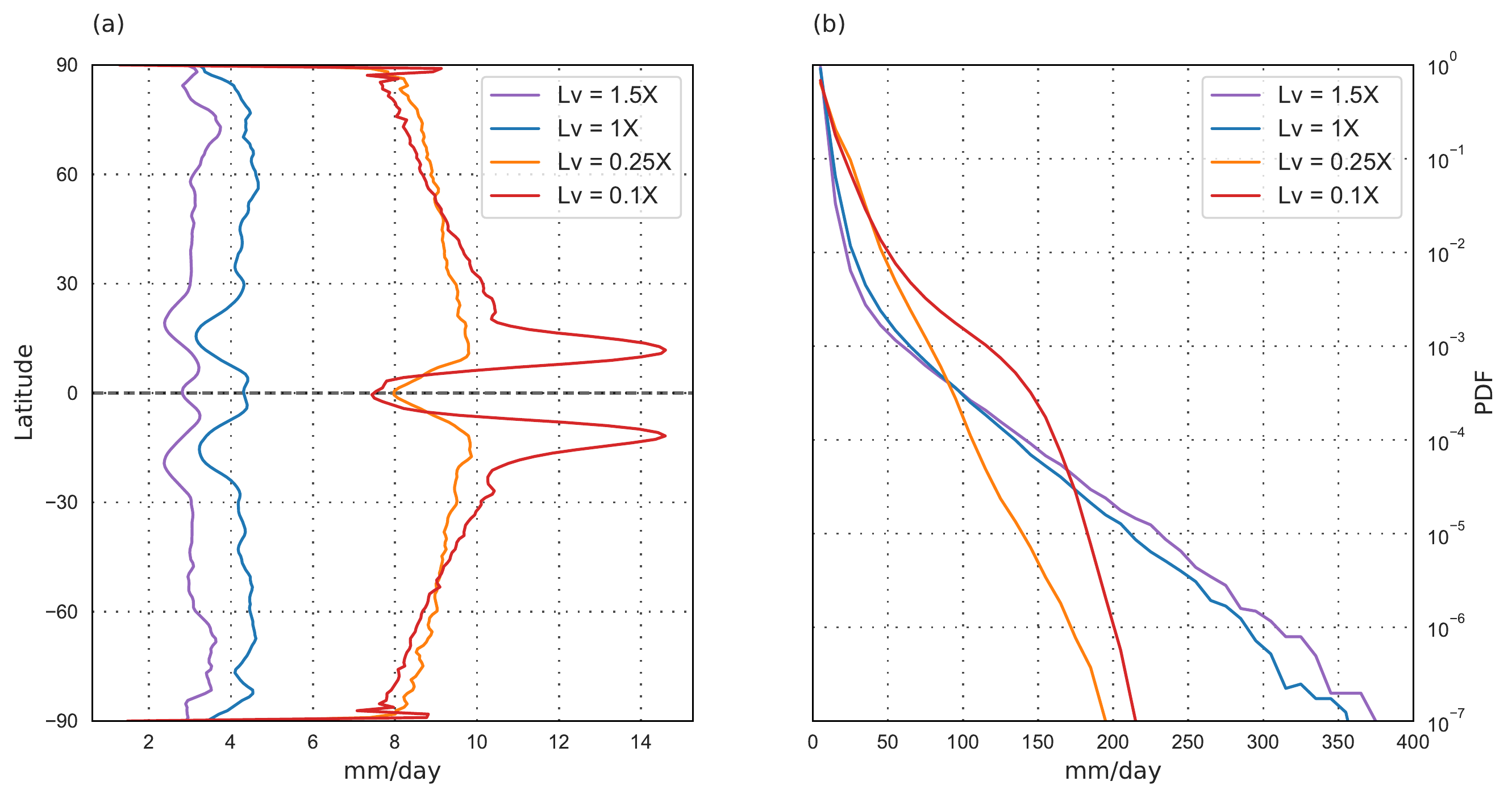}
    \caption{(a) Zonal mean precipitation averaged over the last two years of the run. (b) Probability density function of the daily mean precipitation computed over the the last two years of the run. 
    Four different simulations with $L_v=$ 1.5X, 1X, 0.25X and 0.1X are shown.}
    \label{fig:AquaPrecipitationZonalMeanPdf}
\end{figure}

\noindent Snapshots of the daily precipitation fields are shown in Figure \ref{fig:AquaPrecipitation}. For stronger coupling (say for example, $L_v=$ 1.5X) intense rainfall events (dark colors in Figure \ref{fig:AquaPrecipitation}a) have a very different character in the tropics and midlatitudes. The former have a wave-like nature while the latter are clearly associated with vortices. 
As latent heat decreases, rainfall is more evenly distributed, with relatively higher variability near the tropics and subtropics.  
Further, with weaker coupling, the precipitation has a peak at subtropics (also seen in Figure \ref{fig:AquaPrecipitationZonalMeanPdf}a), corresponding to the ascending motion in the subtropical region (Figure \ref{fig:AquaHadley}). 
However, with stronger coupling the precipitation has an equatorial peak (at least locally), consistent with the ascending motion in the tropics (Figure \ref{fig:AquaHadley}). Thus, the sense of the Hadley cell seen in Figure \ref{fig:AquaHadley}, especially its change in direction of overturning is in line with the zonal mean precipitation. In fact, the near equatorial region gets progressively devoid of strong rainfall events with decreasing $L_v$ which is consistent with the loss of Kelvin and MJO-like modes (Figure \ref{fig:AquaWk}). Rossby waves survive for smaller $L_v$, and are likely responsible for the between gyre or meridionally aligned events in Figure \ref{fig:AquaPrecipitation}c, d \citep{WKW,suhas2020}. Given that there are no meridional gradients at the surface, the organization of moist convection in the tropics by equatorial waves likely give rise to precipitation patterns, which in turn, via latent heating couple to result in large-scale ascent and the overturning flow \citep{horinouchi2012moist}. The connection of tropical wave activity to the Hadley cell is also supported by the fact that the width of these tropical cells, while not an exact match, follows the same pattern as the equatorial deformation scale in Figure \ref{fig:AquaKeSpectra}.

\begin{figure}
    \centering
    \includegraphics[width=\textwidth]{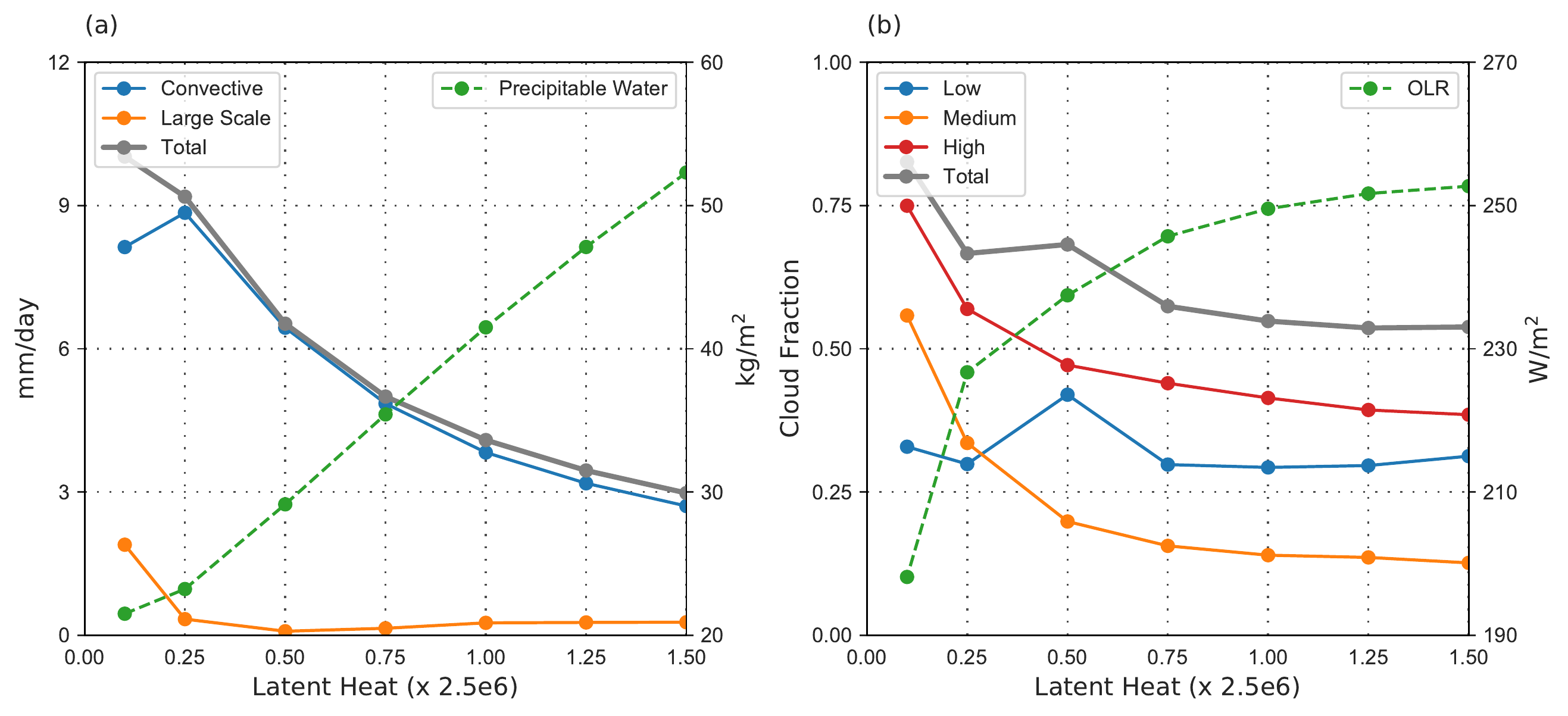}
    \caption{Globally averaged (a) precipitation \& precipitable water and, (b) cloud fraction \& OLR as a  function of latent heat. The fields are averaged over the last two years of the run.} 
    \label{fig:AquaPrecipitationGlobalMean}
\end{figure}

\noindent 
Interestingly, precipitation is actually higher for lower latent heat values at all
latitudes as is seen in Figure \ref{fig:AquaPrecipitationZonalMeanPdf}a. The distribution of these precipitation events (Figure \ref{fig:AquaPrecipitationZonalMeanPdf}b) shows that while the larger latent heat cases have a higher chance of extreme rainfall (up to 400 mm/day), the lower latent heat runs have a higher probability of comparatively moderate rainfall events (about 50-150 mm/day). 
A similar picture emerges in Figure \ref{fig:AquaPrecipitationGlobalMean}a, with the global mean precipitation increasing with decreasing latent heat. Note that precipitable water increases with moist coupling (Figure \ref{fig:AquaPrecipitationGlobalMean}a), perhaps suggesting 
that with decrease in precipitation, more precipitable water is stored in the atmosphere. 
Consistent with the behaviour of precipitation, cloud fraction, shown in Figure \ref{fig:AquaPrecipitationGlobalMean}b, also increases with decreasing $L_v$. This increase is mainly due to changes in the medium and high cloud cover. We also see that outgoing longwave radiation (OLR) saturates at high $L_v$ (Figure \ref{fig:AquaPrecipitationGlobalMean}b), possibly as a result of the increase in water vapor \citep{koll}. Overall, as OLR saturates and the SST is fixed, this results in an increase in tropospheric temperature with height (Figure \ref{fig:AquaTempVerticalProfile}b, c). This implies an increase in stability for larger $L_v$ in the middle and upper troposphere. In turn, this possibly affects middle and high clouds and accounts for the observed decrease in precipitation and cloud fraction \citep{Bony}. 
Interestingly, recent explorations of moist and dry warm core cyclones have also shown an increase in cloud fraction for relatively drier conditions \citep{Chav_dry}.

\section{Conclusion and Discussion}

We have explored the various aspects of dynamically dry and moist atmospheric circulations using a 3--D aquaplanet model with uniform lower boundary conditions. 
The degree or strength of moist coupling was controlled by systematically varying the latent heat of water vapor ($L_v$), and with $L_v \rightarrow 0$, water substance is essentially a passive tracer from a dynamical point of view. This allows one to not only contrast the moist and dry dynamics, but also look at the emergent dynamics in the intermediate moist coupling regimes. In all these experiments, even though the SST is fixed, the atmosphere is in energy balance.

\noindent Despite the absence of meridional thermal gradients, we observe a general circulation with Hadley and Ferrel cells. These cells are of comparable magnitude to those on present-day Earth. But, the nature of the Hadley cell is quite different from an Earth-like scenario where the surface temperature gradient and midlatitude baroclinic waves play a prominent role in determining its terminus \citep{levine-schneider}; in fact, as described by \citet{horinouchi2012moist}, ascent in these cells is possibly dictated by the organization of latent heating via equatorial waves which are prominent with strong coupling.  Interestingly, for weaker moist coupling ($L_v \le $ 0.25X), the mean  circulation reverses with a poleward surface transport in the tropics. In the limiting case of a dry atmosphere, the height of the circulation dramatically reduces, suggesting the importance of moist processes in sustaining the deep convection. Further, tropical cells are thermally indirect and likely also influenced by eddy momentum fluxes. Easterlies form through most of the troposphere with a small region of superrotation in the upper tropics. The magnitude of superrotation weakens and disappears as water vapor becomes passive. Eddy momentum flux convergence is responsible for driving the zonal mean flow in the tropics. In fact, these fluxes are prominent in both the upper and lower troposphere, thus serving as an intermediate regime between the present-day Earth and high obliquity planets with reverse meridional surface temperature gradients, where fluxes are largely confined to the upper and lower troposphere, respectively \citep{ait2015eddy,kang2019tropical}.  
While the upper levels winds in the midlatitudes are mostly westward, westerlies emerge with a weakened moist coupling. This change in direction of the zonal wind is consistent with the sign of the upper tropospheric meridional temperature gradient in each of these runs.

\noindent As expected, the dry static energy flux is dominated by the upper level circulation, while the latent heat flux mimics the sense of mean circulation at the surface. 
In the tropics, MSE transport is determined by the  difference of these two opposing components. 
In all the cases, MSE transport is equatorward in the tropics and poleward in the midlatitudes, with a net poleward transport when averaged over all the latitudes. In addition, transport in the tropics and midlatitudes is largely due to the mean flow and eddy components, respectively. It should be noted that the transport of energy in tropics and midlatitudes is quite different from present-day Earth \citep{ray_deep,Tren_step} in that, in the tropics, the latent flux is larger than the dry term and the net result is an equatorward transport of energy, and in the midlatitudes, both the latent and dry terms are important and directed poleward.
Remarkably the magnitude of moist static energy transport remains roughly invariant from $L_v=$ 1.5X to 0.25X, and then falls drastically as water substance becomes almost passive. 
Much like the observations of \citet{frierson2007gray}, the nearly constant transport is due to an increase of dry static energy transport that compensates the decreasing latent heat component with $L_v$. 
The kinetic energy spectra for the relatively strong coupling runs have some similarities to the Earth's atmosphere; specifically, the power spectrum is characterized by an approximate slope of $-3$ to $-4$ at large scales and is dominated by rotational modes. Though, in contrast to the present day Earth, there are no signs of a transition to a shallower slope at smaller scales. 
With weakening moist coupling, the deformation scale reduces with a corresponding shift in the peak of energy spectra towards the smaller scales and an overall flattening of spectra. In fact, in the passive run, the spectrum suggests a broad peak in variance at about the deformation scale. 

\noindent Transients in the tropics and midlatitudes are also profoundly affected by the moisture interactions. 
For stronger coupling, most of the large-scale variability in the tropics lies along familiar equatorial modes with Earth-like speeds and there is a systematic increase in the time period of these systems with progressively stronger coupling. In particular, as $L_v$ goes from 0.75X to 1.5X, the dominant peak of activity moves from about a month to 90 days. Further, 
intraseasonal modes, especially the eastward moving MJO-like structure and Kelvin waves disappear as water substance becomes dynamically passive. The lack of a MJO-like mode for the passive case suggests the vital and possibly an essential role played by interactive moisture in its dynamics. In accord with the spatial energy spectrum, the passive case does not have much energy at large scales. 
The midlatitudes are characterised by multiple tropical storm-like warm core vortices, drifting poleward and westward over the time period of a month. 
The lack of baroclinic instability due to the absence of imposed meridional thermal gradients, and the conditionally unstable nature of the temperature profiles possibly favour the existence of warm core vortices similar to the Earth-like tropical cyclones even at higher latitudes. These vortices are associated with extreme rainfall events, and are also the reason for a poleward MSE transport in the midlatitudes despite the absence of any imposed temperature surface gradient. As with the ``dry" simulations of tropical cyclones \cite{Chav_dry}, when water substance becomes passive, the storm-like vortices become less intense, smaller in size and their temperature anomalies decrease. The change in the vorticity field is succinctly captured by the shift from a PDF with fat-tails (large $L_v$) to one with a purely exponential form (small $L_v$). The largest temperature anomalies are observed in the upper troposphere, and in addition to the deep storm-like vortices, we note the presence of near tropopause, shallow waves that contribute (albeit a very small amount) to the poleward dry static energy transport. 
The change from wave-like disturbances to warm-core vortices seems to have very strong implications for the zonal mean energy budget and mean temperature structure. While wave disturbances act to mix background gradients diffusively, the moist and warm vortices which move poleward due to a beta drift carry MSE poleward regardless of the direction of the background MSE gradient. As a result, the vortex world has warmer poles and an up gradient MSE flux, while the wave-world has colder poles and down gradient MSE fluxes.

\noindent Snapshots of intense precipitation suggest a change in morphology between tropical and midlatitude events, the former are wave-like while the latter are in the form of vortices. With decreasing $L_v$, the region near the equator is progressively devoid of intense rainfall, this goes hand in hand with the loss of Kelvin and MJO-like modes. Interestingly, global mean precipitation increases with weaker moist coupling with an opposing trend shown by the precipitable water. 
In essence, extreme rainfall events decrease, but there is more rainfall with weaker coupling of water substance. The decrease in rainfall and mid to upper level cloud fraction with increasing $L_v$ appears to be tied to the mechanism proposed by \citet{Bony}, involving increased static stability of the troposphere at middle and upper levels. More fundamentally, OLR saturates as $L_v$ increases, hence with fixed SSTs, the system responds with an increase in tropospheric temperature. This appears to be the cause for an increased static stability and results in lower amounts of total precipitation with strong coupling of water vapor.

\noindent We believe that the results from these simulations, much like the moist and dry scenarios studied by \citet{frierson2006gray,frierson2007gray}, will help in developing a more robust view of planetary atmospheric circulation regimes in the presence of a condensable substance. In particular, the flexibility of our approach could also be of use in probing the dynamics of planetary atmospheres with condensable substances that have a significantly different latent heat than water vapor. From a general circulation point of view, these simulations show a very different way of obtaining tropical and extra-tropical regimes. Specifically, we have a Hadley cell with a well defined terminus but rather than demarcating the change to a baroclinic wave dominated regime \citep{levine-schneider}, here the transition is to storm-like warm core vortices. Indeed, the Hadley cell itself appears to be coupled to the equatorial waves and their modulation of latent heating, rather than a surface temperature gradient. In fact, the width of the Hadley cell is seen to decrease with $L_v$, and this seems to be directly connected to the equatorial deformation scale. 
Further, similar to emerging details on the possibility of warm core cyclones in dry simulations \citep{Ag,Chav_dry,wang_lin}, and their implications for real-world cyclonic systems, the effects of varying moist coupling on transient activity could prove to be of use in a more basic understanding of the moist modes of tropical intraseasonal variability. 


\section*{Acknowledgements}
We would like to thank the Supercomputing Education and Research Centre (SERC) at IISc for computer facilitites where most of these simulations were carried out. 

\bibliographystyle{apalike}
\bibliography{ref.bib}

\end{document}